\documentclass{ptephy_v1}

\preprintnumber{arXiv: 2106.07209 [hep-ph] } 

\begin{document}

\title{Heavy particle non-decoupling in flavor-changing gravitational interactions}


\author[1]{Takeo Inami}
\affil{Theoretical Research Division, Nishina Center,  RIKEN, Wako, 351-0198, Japan \email{inamitakeo@gmail.com}}

\author[2]{Takahiro Kubota}
\affil{CELAS and Department of Physics, Osaka University, Toyonaka, Osaka 560-0043, Japan
\email{takahirokubota859@hotmail.com}}



\begin{abstract}%

The flavor-changing  gravitational  process, 
$d \to s + {\rm graviton}$, 
is  evaluated  at the one-loop level in the standard electroweak theory
with on-shell renormalization.
The results we present in the 't Hooft-Feynman gauge are valid for on- and off-shell quarks and 
for all external and internal  quark masses. 
 We show that there exist non-decoupling effects of the internal heavy top quark 
 in interactions with gravity.  
A naive argument taking account of the quark Yukawa coupling suggests that the amplitude of the process 
$d \to s + {\rm graviton}$ in the large top quark mass limit 
would possibly acquire   an enhancement factor $m_{t}^{2}/M_{W}^{2}$, where $m_{t}$ and  
$M_{W}$ are the top quark and  the $W$-boson masses, respectively.
In practice this leading enhancement is absent in the renormalized amplitude due to cancellation. 
Thus the non-decoupling  of the internal top quark takes place at the ${\cal O}(1)$ level.
The flavor-changing two- and three-point functions are shown to satisfy the  Ward-Takahashi identity, 
 which is used for a consistency-check of  the aforementioned cancellation of the 
${\cal O}(m_{t}^{2}/M_{W}^{2})$ 
terms.  Among the   ${\cal O}(1)$ non-decoupling terms,   we sort out those that can be regarded as 
due to the effective Lagrangian   in which quark bilinear forms are  coupled to the scalar curvature.
\end{abstract}

\subjectindex{B06, B32, B56, B57, E00}

\maketitle

\section{Introduction}
\label{}

The discoveries of the gravitational waves at frequencies $f > 10 \: {\rm Hz} $ by LIGO and Virgo collaboration 
via a binary black hole merger and a binary neutron star inspiral have been hailed as a 
major milestone of gravitational wave astronomy \cite{abbott0, abbott1, abbott2, abbott3}.
The gravitational wave is now expected  to be an exquisite tool not only to study 
astronomical objects such as black holes and neutron stars, 
but also to probe viable extension of general relativity as well as what lies Beyond the Standard Model 
(BSM) of elementary particles. It would be extremely interesting if we could look  into the early Universe 
before the time of last scattering by searching for  gravitational waves.

The recent analyses of the 12.5-year pulsar timing array data at frequencies $ f \sim 1/{\rm yr} $ by NANOGrav Collaboration \cite{nanograv} in search for a stochastic gravitational wave background 
\cite{lommen, tiburzi, burkespolaor}
are also of particular importance 
 and are encouraging enough for us to speculate much about 
BSM: 
 cosmic strings or super-massive black holes  as  possible   sources   of the gravitational wave,    
 first order phase transitions in the dark sector,  new scenarios of leptogenesis induced by gravitational backgrounds  and so on so forth. 
In search for new avenues of BSM with the help of 
stochastic gravitational waves, it would sometimes happen that one has to deal with gravitational 
interactions of heavy unknown particles, in particular, on the quantum level. In such a case we are 
necessarily forced to pay attention to  heavy particle mass effects on physical observables. 

Bearing these new directions in our mind, 
we would like to present in this paper  
  an example in which heavy particles running along 
internal loops in the gravitational backgrounds induce potentially large and important new 
type of interactions.   
Recall that 
an  important issue in particle physics incorporating possible heavy particles has been whether 
heavy particles have power-suppressed and therefore negligiblly small   effects 
in low-energy processes (decoupling), 
or their effects may be  observable in the form of  new induced  interactions in the limit of very large  mass 
(non-decoupling).
To keep our investigation within a reasonable size, we study specifically 
the loop-induced flavor-changing process 
\begin{eqnarray}
d \to s + {\rm graviton} 
\label{eq:dsgraviton}
\end{eqnarray}
in the standard 
electroweak interactions, instead of launching into the BSM studies. 
In our case the top quark is supposed to be the heavy particle as opposed to all the other light quarks. 
The reason for computing (\ref{eq:dsgraviton})
is that the process (\ref{eq:dsgraviton}) is analogous to  
$d \to s + \gamma$
and  $d \to s + {\rm gluon}$ (Penguin) processes 
 and that  the latter two   processes are  known to exhibit top quark non-decoupling effects  
 in   low-energy decay phenomena. 
 It is quite natural to expect that similar non-decoupling phenomena  
 would take place in  (\ref{eq:dsgraviton}) 
 and we will argue  in the present  paper that this expectation 
 is in fact the case.  
So far as we know,  this is the first example  of the  non-decoupling of the internal heavy  top quark  in 
 {\it gravitational interactions of light quarks. }
 
One of the sources   of the non-decoupling   may be  
searched for  in the unphysical scalar field coupling to quarks with the strength proportional 
to the quark masses. This can be seen most apparently in the Feynman rules in the 
't Hooft-Feynman gauge, which we will use throughout. 
We are particularly interested in whether or not the process (\ref{eq:dsgraviton}) would be enhanced by 
the large factor $m_{t}^{2}/M_{W}^{2}$, where $m_{t}$ and $M_{W}$ are the top quark 
and $W$-boson masses, respectively. 
A quick glance over the Feynman rules, in fact, tells us that apparently this large factor comes into 
the Feynman amplitude  as the coupling of exchanged unphysical scalar field to internal 
top quark. However we will show in the present paper by explicit  calculation that  this enhancement 
factor disappears due to 
 cancellation among the terms of  ${\cal O}(m_{t}^{2}/M_{W}^{2})$ in the renormalized 
transition amplitude
\footnote{
When we say ``terms of  ${\cal O}(m_{t}^{2}/M_{W}^{2})$", it is implicitly assumed that logarithmically 
corrected terms such as $(m_{t}^{2}/M_{W}^{2}){\rm log} (m_{t}^{2}/M_{W}^{2})$ are also included. 
Likewise,  ${\cal O}(1)$ terms are assumed to include ${\rm log}(m_{t}^{2}/M_{W}^{2})$ terms as well.
}.
 The breaking of the top quark decoupling thus takes place mildly on the 
${\cal O}(1)$ level.  
We will confirm that  the cancellation of the ${\cal O}(m_{t}^{2}/M_{W}^{2})$ terms is consistent 
with the Ward-Takahashi identity associated with the invariance under the general 
coordinate transformation.

Appelquist and Carazzone \cite{AC}  once pointed out in the mid 1970's that 
virtual  effects of heavy unknown particles
can be safely neglected in  low-energy phenomena,  provided that coupling constants are 
all independent of heavy particle masses. This fact  is often referred to as  the decoupling theorem,  which
provides us with an effective 
strategy to handle  low-energy experimental data without worrying much about 
unknown new physics.  
In the course of the development of particle physics towards 
the end of the last century, however, 
the table has been turned around: we now believe that non-decoupling phenomena 
are much more interesting than  decoupled cases   and that we would perhaps 
be able to have a glimpse  of  high energy contents of the future theory of elementary particles 
by investigating  non-decoupling phenomena.

In the standard electroweak theory, the Higgs boson and unphysical scalar fields are coupled to 
quarks with the strength proportional to the quark masses. For large quark masses as for the case 
of the top quark, the breaking of the decoupling theorem is naturally expected and 
in fact non-decoupling phenomena are ubiquitous in the Standard Model.  They include 
Higgs boson production in the $pp$-collision via gluon fusion process through a  top quark loop 
\cite{georgi, inamikubotaokada, spira}, 
the various decay processes of the Higgs boson involving heavy quarks 
\cite{wilczek,  shifman1, shifman2, inamikubota, sakai, braatenleveille}, 
 heavy quark  effects on the $K_{0}-\overline{K}_{0}$ and $B_{0}-\overline{B}_{0}$  mixings
\cite{gaillardlee, inamilim, buras, deshpande1, deshpande2, botella}, 
etc. 
It is very interesting to see whether a similar explanation for non-decoupling of heavy top 
quark  effects  would work as well  in gravitational interactions of light quarks. This is 
actually a strong motive force for us to examine (\ref{eq:dsgraviton}). 

 After submitting the present  paper for publication,  we  learned
that  the process (\ref{eq:dsgraviton})  had  once  been   computed 
  in the  't Hooft-Feynman gauge and in the unitary gauge by Degrassi et al. \cite{degrasse} 
 and  was  investigated  by Corian{\` o} et al    \cite{coriano1,   coriano2} 
 for a different purpose from ours. Their elaborate calculations, however, 
 are not quite  suitable for our  use   since they put external quarks on the mass shell,  while 
 we would like to make the non-decoupling phenomena manifest by 
 studying    {\it off-shell} effective interactions  in the large top quark mass limit. 
 Our one-loop calculation is made to this end.

The present paper is organized as follows. First of all we explain  in Section \ref{sec:duracfermion}
the method of putting ``weight" on the Fermion fields in the curved background to 
render the Feynman rules to be discussed in Section \ref{sec:electroweaktheory} a little simpler.
The self-energy type $d \to s $ transition in Minkowski space is evaluated in 
Section \ref{sec:selfenergy}, the result of which is closely connected with  the counter terms 
eliminating the divergencies associated with (\ref{eq:dsgraviton}) as argued in 
Section \ref{sec:infinitysubtraction}.  
In Section \ref{sec:gravitationalflavorchangingvertex} we compute 
all the one-loop Feynman diagrams associated with 
 (\ref{eq:dsgraviton}).
 The renormalization constants prepared in Section \ref{sec:infinitysubtraction} are shown in 
Section \ref{sec:cancellation} to be  instrumental to eliminate all the ultraviolet divergences 
in  (\ref{eq:dsgraviton}).  It is argued in Section  \ref{sec:wardtakahashi}  that 
unrenormalized and renormalized quantities associated with  (\ref{eq:dsgraviton}) 
satisfy the same  Ward-Takahashi identity.
The terms in the renormalized transition amplitude behaving 
asymptotically as ${\cal O}(m_{t}^{2}/M_{W}^{2})$ 
in the large top quark mass limit are   investigated  in Section \ref{sec:effectivelagrangian} and are 
shown to vanish via mutual cancellation. The ${\cal O}(1)$   
terms for the large top quark mass are also discussed 
in Section \ref{sec:gravitationalpauliterm}, highlighting those that can be expressed by the operator 
of quark bilinear form coupled to the scalar curvature. 
Section \ref{sec:summary} is devoted to summarizing the present paper.
Various definitions of Feynman parameters' integrations are collected  in  
Appendix \ref{sec:integralrepresentation1} and some combinations thereof are defined in 
Appendix \ref{appendixb}.

\section{Dirac Fermions in gravitational field }
\label{sec:duracfermion}

Techniques of 
loop calculations involving Dirac Fermions in the curved spacetime, which is 
our central concern in studying  (\ref{eq:dsgraviton}), 
were discussed long time ago 
by Delbourgo and Salam  \cite{delbourgosalam, delbourgosalam2} in connection with anomalies 
\cite{kimura, eguchifreund}.  They took a due account of  ``the weight factors" of Fermions 
\cite{ishamsalamstrathdee}, 
which  we now recapitulate while setting up our notations. 
Hereafter in this Section we will use 
Greek  indices $\mu $, $\nu $ etc. for labeling  general coordinates and indices $a$, $b$ etc. 
for labeling  the coordinates in a locally inertial coordinate system. The latter indices 
 are raised and lowered by the Minkowski metric $\eta^{ab}$ and $\eta_{ab} $, respectively.

The Lagrangian of Fermions in the curved spacetime is as usual given by 
\begin{eqnarray}
{\cal L}_{\rm Dirac}
&=&
\sqrt{-g}\left \{
\frac{i}{2}\left (
\overline{\psi}\; \gamma^{\mu} \: \nabla_{\mu}\psi 
- \nabla_{\mu} \overline{\psi} \: \gamma^{\mu} \: \psi 
\right ) -\overline{\psi} \: m\: \psi
\right \}\:,
\label{eq:1}
\end{eqnarray}
where our notations are
\begin{eqnarray}
& &\gamma^{\mu}={e^{\mu}}_{a}\gamma^{a}\:, 
\\
& &
\nabla_{\mu}\psi = \partial_{\mu}\psi  - \frac{i}{4}{\omega_{\mu  ab}}^{}
\sigma^{ab} \: \psi \: , 
\hskip1cm 
\nabla_{\mu}\overline{\psi }= \partial_{\mu}\overline{\psi }
 +  \frac{i}{4} \overline{\psi} \: {\omega_{\mu  ab }}^{}
\sigma^{ab} \: , 
\\
& & \sigma^{ab}=\frac{i}{2}(\gamma^{a}\gamma^{b} - \gamma^{b}\gamma^{a})\: , 
\label{eq:sigmaab}
\end{eqnarray}
and $g={\rm det} \: (g_{\mu \nu})$.
The relation between the spacetime metric $g_{\mu \nu}$ and the vierbein ${e_{\mu}}^{a}$ 
is given as usual  by 
$
g_{\mu \nu} ={e_{\mu}}^{a}\: {e_{\nu}}^{b}\: \eta_{ab}
$.  
The  spin connection $\omega_{\mu  ab}$ is expressed  in terms of the vierbein  as 
\begin{eqnarray}
\omega_{\mu ab}
&=&
\frac{1}{2}\: {e^{\nu}}_{a}\left (
\partial_{\mu}e_{\nu b} - \partial_{\nu}e_{\mu b}
\right )
-
\frac{1}{2}\: {e^{\nu}}_{b}\left (
\partial_{\mu}e_{\nu a} - \partial_{\nu}e_{\mu a}
\right )
-
\frac{1}{2}\: {e^{\rho}}_{a}{e^{\sigma}}_{b}\left (
\partial_{\rho}e_{\sigma c} - \partial_{\sigma }e_{\rho c}
\right ) {e_{\mu}}^{c}\:.
\nonumber \\
\label{eq:spinconnection}
\end{eqnarray}
Noting the identity of gamma matrices 
\begin{eqnarray}
\gamma^{\mu}\sigma^{ab}+\sigma^{ab}\gamma^{\mu}
=
{e^{\mu}}_{c}
\left (
\gamma^{c}\sigma^{ab}+\sigma^{ab}\gamma^{c}
\right )
=-2 {e^{\mu}}_{c}\: \varepsilon^{abcd}\: \gamma_{d}\gamma^{5}\;,
\label{eq:gammamatrixformulae}
\end{eqnarray}
we are able to cast the Dirac Lagrangian (\ref{eq:1}) into 
\begin{eqnarray}
\label{eq:delbourgosalam}
{\cal L}_{\rm Dirac}
&=&
\sqrt{-g}\left \{
\frac{i}{2}\left (
\overline{\psi}\; \gamma^{\mu} \partial_{\mu}\psi -\partial_{\mu} 
\overline{\psi} \gamma^{\mu}\psi 
\right ) -\overline{\psi}m\psi
\right \}
- \frac{1}{4}
 \sqrt{-g} \left (
\overline{\psi}\:  
{e^{\mu}}_{a}
\omega_{\mu bc}\: \varepsilon^{abcd}
\gamma_{d}\:\gamma^{5} 
\psi
\right )\:. 
\nonumber \\
\end{eqnarray}

In order to facilitate perturbative calculations in Section \ref{sec:gravitationalflavorchangingvertex}  
we would like to absorb $\sqrt{-g}$  on the right hand side  of   (\ref{eq:delbourgosalam})
into dynamical fields  as much as possible,   putting  a  weight factor $(-e)^{1/4}$ on the  Dirac fields 
\begin{eqnarray}
\Psi \equiv (-e)^{1/4}\:\psi\: , \hskip1cm
\overline \Psi \equiv (-e)^{1/4}\: \overline \psi\: ,
\label{eq:weighteddirac}
\end{eqnarray}
where 
\begin{eqnarray}
(- e)={\rm det}\: ({e_{\mu}}^{ a})=\sqrt{-g}\;.
\end{eqnarray}
In terms of the weighted Dirac fields (\ref{eq:weighteddirac}), the Dirac Lagrangian (\ref{eq:delbourgosalam}) turns out to be
\begin{eqnarray}
{\cal L}_{\rm Dirac}
&=&
\frac{i}{2}\: {\tilde e^{\mu}}_{\:\:\:a}\left (
\overline{\Psi}\; \gamma^{a} \partial_{\mu}\Psi 
-
\partial_{\mu} 
\overline{\Psi} \gamma^{a}\Psi 
\right ) -
\sqrt{-e}\: \overline{\Psi}m\Psi
- \frac{1}{4}
\overline{
\Psi}\:  
{\tilde e^{\mu}}_{\:\:\:a}\: 
\tilde \omega_{\mu bc}\: \varepsilon^{abcd}
\gamma_{d}\:\gamma^{5} 
\Psi\: .
\nonumber \\
\label{eq:dirac}
\end{eqnarray}
Here we have introduced weighted vierbein
\begin{eqnarray}
\tilde e^{\mu}_{\:\:\: a} = \sqrt{-e}\: {e^{\mu}}_{a}
\label{eq:weightedvierbein}
\end{eqnarray}
and $\tilde \omega_{\mu ab}$ is defined analogously to (\ref{eq:spinconnection}) by 
\begin{eqnarray}
\tilde  \omega_{\mu ab}
&=&
\frac{1}{2}\: {\tilde e^{\nu}}_{\:\:\:a}\left (
\partial_{\mu} \tilde e_{\nu b} - \partial_{\nu} \tilde e_{\mu b}
\right )
-
\frac{1}{2}\: {\tilde e^{\nu}}_{\:\:\:b}\left (
\partial_{\mu} \tilde e_{\nu a} - \partial_{\nu} \tilde e_{\mu a}
\right )
-
\frac{1}{2}\: {\tilde e^{\rho}}_{\:\:\:a}{\tilde e^{\sigma}}_{\:\:\:b}\left (
\partial_{\rho} \tilde e_{\sigma c} - \partial_{\sigma } \tilde e_{\rho c}
\right ){ \tilde e_{\mu}}^{\:\:\: c}\:.
\nonumber \\
\end{eqnarray}
To arrive at (\ref{eq:dirac}), use has been made of an identity
\begin{eqnarray}
{e^{\mu}}_{a}\: \omega_{\mu bc} \: \varepsilon^{abcd}
=
\frac{1}{\sqrt{-e}}\: \tilde e^{\mu}_{\:\:\:a}\: \tilde \omega_{\mu bc}\:\varepsilon^{abcd}\:.
\end{eqnarray}
As we see in (\ref{eq:dirac}), the factor $\sqrt{-e}$ appears only in the mass term.
Also note the relation
\begin{eqnarray}
\sqrt{-e}=\left \{  {\rm det}({\tilde e^{\mu}}_{\:\:a})\right \}^{{\color{black}{1/(D-2)}}}\:,
\label{eq:rootminuse}
\end{eqnarray}
where $D$ is the number of spacetime dimensions. We will use the dimensional method for 
regularization and we do not set $D=4$ .

Putting the weight on the fields as in (\ref{eq:weighteddirac}) and (\ref{eq:weightedvierbein}) 
changes the choice of dynamical variables and 
will lead us to a different set of Feynman rules. It has been known, however, that 
the point transformation of dynamical variables does not alter the structure of S-matrix  
\cite{kamefuchi1, chisholm, borchers}
and therefore we need not worry much about the choice of variables.
In the meanwhile although the weighted field method renders loop calculation a little simpler, 
it hinders us from comparing our calculation directly with the preceding ones 
by Degrassi et al. \cite{degrasse} and by Corian{\` o} et al. \cite {coriano1, coriano2} who 
did not put weight on the vierbein or Fermion fields, either. 

\section{The electroweak theory in the curved background}
\label{sec:electroweaktheory}

We are going to work with the standard $SU(2)_{L} \times U(1)_{Y}$ electroweak theory embedded 
in the curved background field with the metric $g_{\mu \nu}$. Deviation from  the Minkowski 
spacetime  is described, in terms of the vierbein,  as
\begin{eqnarray}
{{\tilde e}^{\mu}}_{\:\:\:a}={\eta^{\mu}}_{a} + \kappa \: {h^{\mu}}_{a}\:, 
\label{eq:expansioninkappa}
\end{eqnarray}
where $\kappa =\sqrt{8\pi G}$, $G$ being the Newton's constant. 
In terms of the metric, fluctuations are expressed as 
\begin{eqnarray}
\tilde g^{\mu \nu}={{\tilde e}^{\mu}}_{\:\:\:a}{{\tilde e}^{\nu}}_{\:\:\:b}\: \eta^{ab}
=
\eta^{\mu \nu} + \kappa\: \left ( h^{\mu \nu} + h^{\nu \mu} \right ) 
+ \kappa^{2} h^{\mu \lambda}{h^{\nu}}_{\lambda}\: , 
\end{eqnarray}
where Greek and Latin indices are no more distinguished and indices of $h^{\mu \nu}$ are
raised and lowered by the Minkowski metric. Also from here we assume that $h^{\mu \nu}$ 
is symmetric i.e., $h^{\mu \nu}=h^{\nu \mu}$. 
Also note that (\ref{eq:rootminuse}) gives rise to the formula 
\begin{eqnarray}
\sqrt{-e}=1+\frac{\kappa}{
{\color{black}{D-2}}
}{\eta_{\mu }}^{a}{h^{\mu}}_{a}+ \cdots \cdots \:\: .
\label{eq:rootminuse2}
\end{eqnarray}

In the  $R_{\xi}$-gauge
in the curved background,  we add the following gauge-fixing terms to the action 
\begin{eqnarray}
{\cal L}_{\rm g.f.}
&=&
-
\frac{1}{\xi }\sqrt{-g} \: 
\Big \vert   g^{\lambda \rho} \nabla_{\lambda}W_{\rho} 
{\color{black}{-}} 
i\: \xi \; M_{W}\: \chi 
\Big \vert ^{2}
-\frac{1}{2 \xi ^{\prime}}
\sqrt{-g}\left (
g^{\mu \nu} \: \nabla_{\mu}Z_{\nu} +\xi ^{\prime} \; M_{Z} \:  \chi_{0} \right )^{2}
\nonumber \\
& & 
-\frac{1}{2 \alpha}\: \sqrt{-g}\; 
\left (
g^{\mu \nu}\: \nabla_{\mu}A_{\nu}
\right )^{2}\: , 
\label{eq:gaugefixing}
\end{eqnarray}
where $\xi$, $\xi^{\prime}$ and $\alpha$ are gauge parameters. The masses of $W$- and $Z$-bosons 
are denoted by $M_{W}$ and $M_{Z}$, respectively. 
The electromagnetic field is denoted by $A_{\mu}$  and 
$\chi $ and $\chi _{0}$ are charged and neutral unphysical scalar fields, respectively. 
In our actual calculations we will use the 
$\xi =1$ 't Hooft-Feynman gauge,  in which the $W$-boson propagator is very much simplified and   
is most convenient to deal with.
The second and third terms in (\ref{eq:gaugefixing}) are not  relevant to our later calculations but are 
included here just for completeness. The gravitational field is  an external field and therefore 
the general covariance is not gauge-fixed.

The electroweak Lagrangian in the curved space is given in the power-series  expansion in $\kappa $, 
namely, 
\begin{eqnarray}
{\cal L}&=&{\cal L}_{\rm SM} + \kappa \: h^{\mu \nu}T_{\mu \nu}+ {\cal O}(\kappa^{2})\: , 
\label{eq:startingpointlagrangian}
\\
T_{\mu \nu}&=&T_{\mu \nu}^{(W)}+T_{\mu \nu}^{(\chi)}+ \sum_{q} T_{\mu \nu}^{(\rm q)} 
+ T_{\mu \nu}^{\rm (qW)} + T_{\mu \nu}^{{\rm (q}\chi )}  \:, 
\label{eq:energymomentumtensor1}
\end{eqnarray}
where ${\cal L}_{\rm SM}$ is the standard electroweak Lagrangian in the flat Minkowski space 
and the second term in the expansion  in $\kappa $  in (\ref{eq:startingpointlagrangian}) corresponds to  the one-graviton emission. 
Since we will not consider graviton loops,  the Einstein-Hilbert  gravitational action is not 
included in  (\ref{eq:startingpointlagrangian}) .
The summation in the third term of (\ref{eq:energymomentumtensor1}) is taken over  all quark flavors 
$(q=u, d, s, \cdots )$
and each term in (\ref{eq:energymomentumtensor1}) is respectively given  by
\begin{eqnarray}
T^{(W)}_{\mu \nu}
&=& {{V^{}}_{\mu \nu}}^{\sigma \tau  \lambda \rho}\:\:
( \partial _{\sigma}W_{\tau}^{\dag}) 
\: ( \partial_{\lambda}W_{\rho}) 
+ 2\:M_{W}^{2} \:{\eta^{\tau}}_{( \mu}  {\eta_{\: \nu )}}^{\rho}
 \: W_{\tau}^{\dag}W_{\rho}   
\nonumber \\
& & 
+\frac{2}{\xi }\: {\eta^{\sigma}}_{( \mu} \: {\eta_{\nu )}}^{\tau} \: \eta^{\lambda \rho} 
\:  \left (
W^{\dag}_{\tau} \: \partial_{\sigma} \partial_{\lambda} W_{\rho}
\right )
+\frac{2}{\xi }\: {\eta^{\sigma}}_{( \mu} \: {\eta_{\nu )}}^{\rho} \: \eta^{\lambda \tau} 
\: \left ( W_{\rho}\:
 \partial_{\lambda} \partial_{\sigma} W^{\dag}_{\tau}
\right )\:, 
 \label{eq:twmunu}
\\
T^{(\chi)}_{\mu \nu}&=& 
\partial_{\mu}\chi^{\dag}\: \partial _{\nu}\chi + \partial_{\nu}\chi^{\dag}\: \partial _{\mu}\chi 
- \eta_{\mu \nu}\: 
{\color{black}{\frac{2}{D-2}}}
\: \xi \:M_{W}^{2}\chi^{\dag}\chi  \:,
\\
T_{\mu \nu}^{\rm (q)}&=&\frac{i}{2}\:
\overline{\Psi}_{q} \left (\gamma_{\mu} \overleftrightarrow{\partial_{\nu} }
+
\gamma_{\nu} \overleftrightarrow{\partial_{\mu} }
\right )\:\Psi_{q} -
{\color{black}{\frac{1}{D-2}}}
 \: \eta_{\mu \nu} \overline{\Psi}_{q}\:m_{q} \Psi_{q} \: ,
\label{eq:tqmunu}
\\
T^{\rm(qW)}_{\mu \nu}
&=&
- \: \frac{g}{2 \sqrt{2}\: }\left (
\overline{U_{L}}\gamma_{\mu} V_{\rm CKM}D_{L}W^{\dag}_{\nu}
+
\overline{U_{L}}\gamma_{\nu} V_{\rm CKM}D_{L}W^{\dag}_{\mu}
\right ) + ({\rm h.c.})\:,
\label{eq:tqwmunu}
\\
T^{{\rm (q}\chi)}_{\mu \nu}&=&
 \eta_{\mu \nu}\: 
 {\color{black}{\frac{1}{D-2}}}
 \frac{\sqrt{2}}{v}\:\chi^{\dag} \: \left (
\overline{U_{R}}\:{\cal M}_{u}\:V_{\rm CKM}^{}\:D_{L}
-
\overline{U_{L}}\:V_{\rm CKM}{\cal M}_{d}\:D_{R}
\right ) +({\rm h.c.})\:.
\label{eq:tqchimunu}
\end{eqnarray}
The  quantity ${{V^{}}_{\mu \nu}}^{\sigma \tau  \lambda \rho}$ in (\ref{eq:twmunu})
 is defined by 
\begin{eqnarray}
{{V^{}}_{\mu \nu}}^{\sigma \tau  \lambda \rho}
&=&
-2\: {\eta^{\sigma}}_{( \mu}\: {\eta_{\nu )}}^{\lambda}\:  \eta^{\tau \rho} 
-2
\: {\eta^{\tau}}_{( \mu} \: {\eta_{\nu)}}^{\rho} \: \eta^{\sigma \lambda}
+ 2\: {\eta^{\tau}}_{(\mu}  \: {\eta_{\nu)}}^{\lambda} \: \eta^{\sigma \rho} 
+ 2
 \: {\eta^{\sigma}}_{( \mu} \: {\eta_{\nu ) }}^{\rho} \: \eta^{\tau \lambda} 
\nonumber \\
& & 
+ 
{\color{black}{\frac{2}{D-2}}}\: \eta_{\mu \nu}\:  \left ( \eta^{\sigma \lambda}\:  \eta^{\tau \rho} 
- \: \eta^{\tau \lambda} \: \eta^{\sigma \rho} 
 + \frac{1}{\xi}\: \eta^{\sigma \tau} \: \eta^{\lambda \rho}   
\right )\;,
\label{eq:6indexedV}
\end{eqnarray}
and the symmetrization with respect to indices in a pair of parentheses in (\ref{eq:6indexedV}) is 
done in the following manner
\begin{eqnarray}
A_{(\sigma}B_{\tau)} \equiv \frac{1}{2}\left ( A_{\sigma}B_{\tau}+A_{\tau}B_{\sigma}  \right )\:.
\end{eqnarray}
The symbol of left-right derivative in (\ref{eq:tqmunu}) is defined by 
\begin{eqnarray}
\overleftrightarrow{\partial_{\mu}}=\frac{1}{2}\left (
\overrightarrow{\partial_{\mu}}
-\overleftarrow{\partial_{\mu}}
\right )\:.
\end{eqnarray}
The Cabibbo-Kobayashi-Maskawa (CKM) matrix is denoted by $V_{\rm CKM}$ in 
(\ref{eq:tqwmunu}) and (\ref{eq:tqchimunu}) and 
the diagonal mass matrices of up- and down-type quarks are given, respectively, by
\begin{eqnarray}
{\cal M}_{u}=\left (
\begin{tabular}{ccc}
$m_{u}$ & 0 & 0
\\
0 & $m_{c}$ & 0
\\
0 & 0 & $m_{t}$
\end{tabular}
\right )\:, 
\hskip1cm
{\cal M}_{d}=\left (
\begin{tabular}{ccc}
$m_{d}$ & 0 & 0 
\\
0 & $m_{s}$ & 0
\\
0 & 0 & $m_{b}$
\end{tabular}
\right )\: .
\end{eqnarray}

\vskip0.2cm
The left ($L$)- and right ($R$)-handed  quarks are projected as usual by 
\begin{eqnarray}
L=\frac{1-\gamma^{5}}{2}, \hskip1cm R=\frac{1+\gamma^{5}}{2}, 
\end{eqnarray}
and the projected  up- and down-type quarks are expressed as  
\begin{eqnarray}
U_{L}
=
L\left (
\begin{tabular}{c}
$\Psi_{u}$
\\
$\Psi_{c}$
\\
$\Psi_{t}$
\end{tabular}
\right )_{}
\:,
\hskip1.2cm
D_{L}=
L
\left (
\begin{tabular}{c}
$\Psi_{d}$
\\
$\Psi_{s}$
\\
$\Psi_{b}$
\end{tabular}
\right )_{}
\:,
\\
U_{R}
=
R\left (
\begin{tabular}{c}
$\Psi_{u}$
\\
$\Psi_{c}$
\\
$\Psi_{t}$
\end{tabular}
\right )_{}
\:,
\hskip1.2cm
D_{R}=
R
\left (
\begin{tabular}{c}
$\Psi_{d}$
\\
$\Psi_{s}$
\\
$\Psi_{b}$
\end{tabular}
\right )_{}
\:,
\end{eqnarray}
in (\ref{eq:tqwmunu}) and (\ref{eq:tqchimunu}) . 
The $SU(2)_{L}$ gauge coupling is denoted by $g$ in (\ref{eq:tqwmunu})  and $v$ in 
(\ref{eq:tqchimunu}) is the vacuum expectation 
value.

Before closing this section,  let us add a comment on the relation between the energy-momentum 
tensor and our $T_{\mu \nu}$. The conventional energy-momentum tensor is defined as 
the functional derivative  of the action 
under the variation $ \delta e^{\mu a}$, while $T_{\mu \nu}$ of (\ref{eq:energymomentumtensor1})
is the functional derivative of the action under $\delta {\tilde e}^{\mu a}$. 
The connection between the two types of functional derivatives is given by 
\begin{eqnarray}
\frac{\delta }{\delta e^{\: \mu  a}}
=
\sqrt{-e}\frac{\delta }{\delta {\tilde e}^{\: \mu  a}}
-\frac{1}{2}\sqrt{-e}e_{\:\mu a} \: e^{\: \lambda b}\frac{\delta }{\delta {\tilde e}^{\: \lambda   b}}\;, 
\end{eqnarray}
and therefore the conserved energy-momentum tensor in the flat-space limit is a linear combination 
of $T_{\mu \nu}$ given by
\begin{eqnarray}
T_{\mu \nu}-\frac{1}{2}\:\eta _{\mu \nu} \: \eta^{\lambda \rho}\: T_{\lambda \rho}\:.
\label{eq:conservedenergymomentumtensor}
\end{eqnarray}
The Ward-Takahashi identity which will be discussed later in Section \ref{sec:wardtakahashi} 
is associated with 
(\ref{eq:conservedenergymomentumtensor}).
\section{Self-energy type $d \to s$ transition}
\label{sec:selfenergy}

The purpose of the present work is to uncover the non-decoupling  nature of the internal 
heavy top quark  in the low-energy process (\ref{eq:dsgraviton}), which is induced at the loop levels.  
There are eight one-loop diagrams of two different types, which   will be shown later in 
 Section \ref{sec:gravitationalflavorchangingvertex}
(i.e., Figure \ref{fig:attachedtovertex} and Figure \ref{fig:attachedtopropagators}).  
There the gravitons ($h_{\mu \nu}$) are expressed by  double-wavy  lines 
and are  attached to vertices in Figure \ref{fig:attachedtovertex} and to internal propagators 
in  Figure \ref{fig:attachedtopropagators}. 
The internal quark propagator consists of 
$j=$ top ($t$), charm ($c$) and up ($u$) quarks and we are interested 
in the large top quark mass behavior 
of the amplitudes of Figure \ref{fig:attachedtovertex} and Figure \ref{fig:attachedtopropagators}.

These one-loop contributions to  (\ref{eq:dsgraviton}) 
will be computed in Section  \ref{sec:gravitationalflavorchangingvertex}
and they turn out to be ultraviolet divergent.
The divergences should be subtracted  by using the corresponding counter term Lagrangian 
$\widehat{\cal L}_{\rm c.t.}$  in the curved spacetime, 
 which should be diffeomorphism invariant  and will be given 
in  Section \ref{sec:infinitysubtraction}. 
   Diagrammatically the counter  term
in $\widehat{\cal L}_{\rm c.t.}$ with one external graviton  
will be  denoted by a cross in Figure \ref{fig:counterterm} (b).  
As it turns out,  the flat spacetime limit   ${\cal L}_{\rm c.t.}$   of $\widehat{\cal L}_{\rm c.t.}$ 
should serve as the counter term Lagrangian that is supposed to eliminate
 the divergences associated with the self-energy type 
 $d \to s$ transition $\Sigma (p) $ in the  Minkowski space (without graviton emission). 
 The vertex associated with  ${\cal L}_{\rm c.t.}$   will be    denoted  by a cross in 
 Figure \ref{fig:counterterm} (a).  
 
\begin{figure}[hbt]
\begin{center}
\input{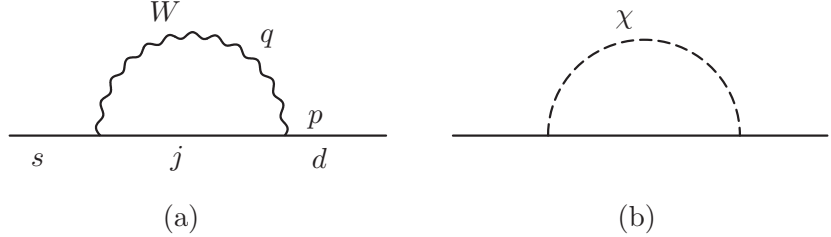}
\vskip0.3cm
\caption{The self-energy type $d \to s$ transition via (a) $W$- and (b) charged unphysical 
scalar  ($\chi $) - boson  exchanges. The intermediate quark is denoted by $j$ ($j=t, c, u$).}
\label{fig:exercise2}
\end{center}
\end{figure}

Now by turning the other way  reversed, we may proceed along  the following way: 
namely,   after computing   $\Sigma (p)$ in this Section \ref{sec:selfenergy}, 
we first   work out   in Section \ref{subsec:countertermsintheflatspacetime}
the renormalization constants contained in  ${\cal L}_{\rm c.t.}$, 
and then  we   deduce  in Section \ref{subsec:countertermsinthecurvedspacetime} 
an explicit  form of $\widehat{\cal L}_{\rm c.t.}$ by employing the diffeomorphism 
invariance argument. 
We will confirm in Section  \ref{sec:cancellation}  that our $\widehat{\cal L}_{\rm c.t.}$ thus obtained is 
necessary and sufficient  to eliminate all the divergences that appear in the one-loop induced $d$-$s$-graviton vertex 
to be computed  in Section  \ref{sec:gravitationalflavorchangingvertex}.
 Keeping these procedures in our mind,  
we would like to start  with the calculation of   
$\Sigma (p)$  
in the flat Minkowski space. 
The $d \to s $ transition takes place at the one-loop level
via $W$-and charged unphysical scalar boson ($\chi$) exchanges  
as  depicted   in Figure  \ref{fig:exercise2}.  
The Feynman rules in the 't Hooft-Feynman gauge ($\xi =1$)   lead us to 
\begin{eqnarray}
 \Sigma (p)
& =  &\sum_{j=t,c,u} (V_{\rm CKM})_{js}^{*} (V_{\rm CKM})_{jd} \left \{ 
  {\cal S}^{(a)}(p)    + {\cal S} ^{(b)}(p) 
 \right \}\:,
 \label{eq:selfenergy}
 \end{eqnarray}
 where we have defined   the integrations of the following forms
\begin{eqnarray}
 {\cal S} ^{(a)} (p)
& \equiv & -i 
\left ( \frac{-ig}{\sqrt{2}} \right )^{2} 
\mu^{4-D}
\int \frac{d^{D}q}{(2\pi)^{D}} \: \gamma_{\alpha}L\frac{i}{\gamma \cdot (p-q) - m_{j}} 
\: \gamma_{\beta} L
\frac{-i \: \eta ^{\alpha \beta} }{q^{2}-M_{W}^{2}}\:,
\label{eq:ia1first}
\\
  {\cal S} ^{(b)}  (p)
& \equiv & -i 
\left ( \frac{- i g}{\sqrt{2}} \right )^{2} \frac{\mu^{4-D}}{M_{W}^{2}}
\int \frac{d^{D}q}{(2\pi)^{D}} 
\left ( m_{j}R - m_{s}L \right ) \frac{i}{\gamma \cdot (p-q) - m_{j}} \left ( m_{j}L -m_{d}R \right )
\nonumber \\
& &  \hskip3cm \times 
 \frac{i}{q^{2}-  M_{W}^{2}}\:, 
 \label{eq:ibfirst}
\end{eqnarray}
in correspondence  to Figure   \ref{fig:exercise2} (a) and Figure   \ref{fig:exercise2}  (b), respectively.
Here $\mu$ is the mass scale of the $D$-dimensional regularization method.

The integrations in (\ref{eq:ia1first}) and (\ref{eq:ibfirst}) are rather standard and we find 
\begin{eqnarray}
{\cal S}^{(a)}(p)
&=&
\frac{-g^{2}}{(4\pi)^{2}}
\left  [    \frac{1}{D-4} + \frac{1}{2} 
+\:f_{1}( p^{2} ) \right ] \gamma \cdot p\: L\:, 
\label{eq:ia1}
\\
{\cal S}^{(b)}(p)
&=&
\frac{-g^{2}}{(4\pi)^{2}}\bigg [
\frac{1}{D-4} \left \{
\frac{1}{2M_{W}^{2}}\:\gamma \cdot p \: (m_{j}^{2}\: L  + m_{s}m_{d}R)
-\frac{m_{j}^{2}}{M_{W}^{2}}(m_{s}L + m_{d}R)
\right \}
\nonumber \\
& & +\frac{1}{2M_{W}^{2} }\left \{
f_{1}(p^{2})\: \gamma \cdot p\:(m_{j}^{2}\: L + m_{s}m_{d}\:   R)
- f_{2}(p^{2})\:m_{j}^{2} (m_{s}L + m_{d}R ) \right \} \bigg ]\:, 
\nonumber \\
\label{eq:sb}
\end{eqnarray}
where $f_{1}(p^{2})$ and $f_{2}(p^{2})$ are defined respectively by 
(\ref{eq:f1(pp)}) and (\ref{eq:f2(pp)}) in Appendix \ref{sec:integralrepresentation1}\:.
Note that  terms  independent of $m_{j}$ that are present in 
  ${\cal S}^{(a)}$ and ${\cal S}^{(b)}$  
will disappear after the $j$-summation in  (\ref{eq:selfenergy}) because of the unitarity   of the 
CKM  matrix,  $V_{\rm CKM}$, i.e.,
\begin{eqnarray}
\sum_{j=t,c,u}  (V_{\rm CKM})_{js}^{*}(V_{\rm CKM})_{jd} =0\:.
\label{eq:ckmunitarity}
\end{eqnarray} 
Also remember that both of  $f_{1}(p^{2})$ and $f_{2}(p^{2})$  contain $m_{j}^{2}$ in their definitions, 
(\ref{eq:f1(pp)}) and (\ref{eq:f2(pp)}).
Therefore putting  (\ref{eq:ia1}) and (\ref{eq:sb}) together, we end up with the formula of 
the self-energy type $d \to s$ transition 
\begin{eqnarray}
\Sigma (p)
&=&
\frac{-g^{2}}{(4\pi )^{2}}
\sum_{j=t,c,u}(V_{\rm CKM})^{*}_{js}(V_{\rm CKM})_{jd}
\:\: \bigg [ f_{1}(p^{2})\;\gamma \cdot p\:L
\nonumber \\
& & 
+ \frac{1}{D-4} \left \{
\frac{m_{j}^{2}}{2M_{W}^{2}}\:\gamma \cdot p \: L  
-\frac{m_{j}^{2}}{M_{W}^{2}}(m_{s}L + m_{d}R)
\right \}
\nonumber \\
& & +\frac{1}{2M_{W}^{2} }\left \{
f_{1}(p^{2}) \: \gamma \cdot p \left ( m_{j}^{2} \: L + m_{s}m_{d} \: R \right )
- f_{2} \left ( p^{2})\:m_{j}^{2} (m_{s}L + m_{d}R \right ) \right \} 
\bigg ]\:.
\nonumber \\
\label{eq:unrenormalisedsigma(p)}
\end{eqnarray}

\section{Infinity subtraction procedure}
\label{sec:infinitysubtraction}

\subsection{Counter terms in the flat-spacetime }
\label{subsec:countertermsintheflatspacetime}

Let us now move to the  subtraction of infinities from $\Sigma (p)$, taking into account  the 
counter terms which are of  the following form
\begin{eqnarray}
{\cal L}_{\rm c.t.}
&=&
Z_{L} \: \overline{\psi_{s}}_{L}\:  i  \gamma \cdot \overleftrightarrow{\partial}   \: \psi_{d \: L}
+
Z_{R} \: \overline{\psi_{s}}_{R} \:  i  \gamma \cdot \overleftrightarrow{\partial}   \: \psi_{d\: R}
\nonumber \\
& & +
Z_{Y1}\: \overline{\psi_{s}}_{R} m_{s} \psi _{d\: L}
+
Z_{Y2}\: \overline{\psi_{s}}_{L}\: m_{d} \psi_{d\: R}
\nonumber \\
& & +({\rm h.c.})\:.
\label{eq:counterterm}
\end{eqnarray}
Here the wave-function renormalization constants,  $Z_{L}$, $Z_{R}$, $Z_{Y1}$ and $Z_{Y2}$
take care of the mixing between $d$- and $s$-quarks under renormalization.
\vskip0.5cm
\begin{figure}[hbt]
\begin{center}
\input{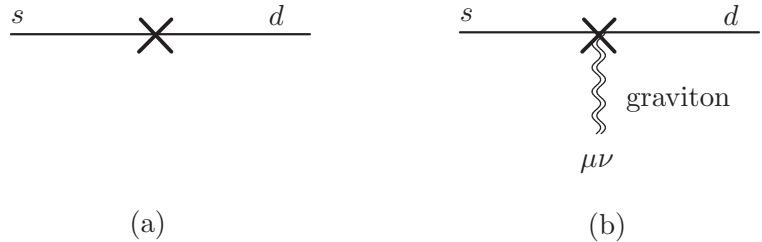}
\caption{The counter term diagram of (a) $d \to s $ transition and (b) $d$-$s$-graviton vertex. 
The insertion of counter terms is indicated by a cross and 
the double wavy line in (b) denotes   an emitted  graviton ($h_{\mu \nu}$).}
\label{fig:counterterm}
\end{center}
\end{figure}

The contribution of 
(\ref{eq:counterterm}) to $d \to s $ transition 
is depicted in   Figure  \ref{fig:counterterm} (a) and is written as  
\begin{eqnarray}
\Sigma _{\rm c.t.} (p)
&=&
Z_{L} \gamma \cdot p \;L + Z_{R}\gamma \cdot p \:R + Z_{Y1}m_{s}L + Z_{Y2}m_{d}R\; .
\label{eq:selfenergycounterterm}
\end{eqnarray}
The renormalization constants are  arranged so that the renormalized 
$d \to s$ transition amplitude 
\begin{eqnarray}
\Sigma_{\rm ren}(p)=\Sigma_{}(p) + \Sigma_{\rm c.t.}(p)
\label{eq:renormalizedselfenergydefined}
\end{eqnarray}
is finite. In other words the renormalization constants are given  the following  form, 
\begin{eqnarray}
Z_{L}
&=&
\frac{g^{2}}{(4\pi)^{2}}
\sum_{j=t,c,u}(V_{\rm CKM})^{*}_{js}(V_{\rm CKM})_{jd}
\:\:  \left \{
\frac{m_{j}^{2}}{2M_{W}^{2}} \cdot  \frac{1}{D-4} \:
- c_{1}(m_{j}) 
\right \}\;,
\label{eq:zl}
\\
Z_{R}
&=&
\frac{g^{2}}{(4\pi)^{2}}
\sum_{j=t,c,u}(V_{\rm CKM})^{*}_{js}(V_{\rm CKM})_{jd}
\:\: \times \left \{-c_{2}(m_{j}) \right \}\;,
\label{eq:zr}
\\
Z_{Y1}
&=&
\frac{g^{2}}{(4\pi)^{2}}
\sum_{j=t,c,u}(V_{\rm CKM})^{*}_{js}(V_{\rm CKM})_{jd}
\:\:  \left \{
-  \frac{m_{j}^{2}}{M_{W}^{2}}\cdot  \frac{1}{D-4} \:
- c_{3}(m_{j})  
\right \}\;,
\label{eq:zy1}
\\
Z_{Y2}
&=&
\frac{g^{2}}{(4\pi)^{2}}
\sum_{j=t,c,u}(V_{\rm CKM})^{*}_{js}(V_{\rm CKM})_{jd}
\:\:  \left \{
- \frac{m_{j}^{2}}{M_{W}^{2}}\cdot  \frac{1}{D-4} \: 
- c_{4}(m_{j})  
\right \}\;, 
\label{eq:zy2}
\end{eqnarray}
in order to subtract the  $D=4$ pole terms in $\Sigma (p)$. 
Here $c_{1}(m_{j}) $, $c_{2}(m_{j}) $, $c_{3}(m_{j}) $ and $c_{4}(m_{j}) $ are 
all finite  and should be determined  by 
specifying  the subtraction conditions.

Now we adopt the on-shell subtraction conditions   \cite{deshpande1}
 in such a way that the renormalized self-energy
$\Sigma_{\rm ren}(p)$ should satisfy the following conditions
\begin{eqnarray}
& & \Sigma_{\rm ren}\:\Psi_{d}=0, \hskip1cm {\rm for}\:\: p^{2}=m_{d}^{2}\:,
\nonumber \\
& & \overline{\Psi}_{s}\:\Sigma _{\rm ren}=0\:,  \hskip1cm {\rm for}\:\: p^{2}=m_{s}^{2}\:.
\label{eq:subtractioncondition}
\end{eqnarray}
Each of the conditions  in (\ref{eq:subtractioncondition}) gives rise to two constraints 
on $\Sigma_{\rm ren}$ : 
one for  left-handed part  and the other for right-handed part. We have   therefore 
  four constraints in total 
in (\ref{eq:subtractioncondition}) 
which in turn determine the four constants $c_{1}(m_{j})  $, $c_{2}(m_{j})  $, $c_{3}(m_{j})  $ and 
$c_{4}(m_{j})  $.

In order to determine these constants on the basis of  (\ref{eq:subtractioncondition}), 
let us note that  (\ref{eq:renormalizedselfenergydefined}) is written  explicitly as 
\begin{eqnarray}
\Sigma _{\rm ren}(p)
&=&
\frac{-g^{2}}{(4\pi)^{2}}
\sum_{j=t,c,u}(V_{\rm CKM})^{*}_{js}(V_{\rm CKM})_{jd}
\:\: \bigg [
\nonumber \\
& & \hskip-1cm
\left \{
c_{1}(m_{j})  +\:  f_{1}(p^{2}) \left ( 1+ \frac{m_{j}^{2}}{2M_{W}^{2}}  \right )
\right \}\: \gamma \cdot p \: L
+ \left \{
c_{2}(m_{j})  + \frac{m_{s} \: m_{d}}{2M_{W}^{2} } \: f_{1}(p^{2}) \right \} \;\gamma \cdot p \: R
\nonumber \\
& & 
\hskip-1cm
+ \left \{
c_{3}(m_{j})  - \frac{m_{j}^{2}}{2M_{W}^{2}}\: f_{2}(p^{2}) \: \right \}  m_{s}  L
 + \left \{
c_{4}(m_{j})  - \frac{m_{j}^{2}}{2M_{W}^{2}} f_{2}(p^{2}) \: \right \}  m_{d}  R  \bigg ]\;.
\label{eq:renormaliseddstransitionamplitudes}
\end{eqnarray}
The subtraction conditions (\ref{eq:subtractioncondition})  then turn out to be 
\begin{eqnarray}
\left \{
c_{1}(m_{j}) + f_{1}(m_{d}^{2}) \left ( 1+\frac{m_{j}^{2}}{2M_{W}^{2} } \right ) \right \} m_{d} 
+
 \left \{
c_{4}(m_{j})  - \frac{m_{j}^{2}}{2M_{W}^{2} } \: f_{2}(m_{d}^{2})
\right \}m_{d}&=&0\:,
\label{eq:subtractioncondition1}
\\
\left \{
c_{2}(m_{j})   + \frac{m_{s}m_{d}}{2M_{W}^{2}} \: f_{1}(m_{d}^{2}) \right \} m_{d} + \left \{
c_{3}(m_{j})   - \frac{m_{j}^{2}}{2M_{W}^{2}}\:f_{2}(m_{d}^{2}) 
\right \} m_{s} &=&0\:, 
\label{eq:subtractioncondition2}
\\
\left \{
c_{1}(m_{j})   + f_{1}(m_{s}^{2}) \left ( 1+\frac{m_{j}^{2}}{2M_{W}^{2} } \right ) \right \} m_{s} 
+
 \left \{
c_{3}(m_{j})   - \frac{m_{j}^{2}}{2M_{W}^{2} } \: f_{2}(m_{s}^{2})
\right \}m_{s}&=&0\:,
\label{eq:subtractioncondition3}
\\
\left \{
c_{2}(m_{j})  + \frac{m_{s}m_{d}}{2M_{W}^{2}} \: f_{1}(m_{s}^{2}) \right \} m_{s} + \left \{
c_{4}(m_{j})   - \frac{m_{j}^{2}}{2M_{W}^{2}}\:f_{2}(m_{s}^{2}) 
\right \} m_{d} &=&0\:, 
\label{eq:subtractioncondition4}
\end{eqnarray}
and we have worked out  the following solutions to 
Eqs. (\ref{eq:subtractioncondition1})-(\ref{eq:subtractioncondition4}), 
\begin{eqnarray}
c_{1}(m_{j}) &=&
\frac{1}{m_{d}^{2}-m_{s}^{2} } \bigg [
-\left \{
m_{d}^{2} f_{1}(m_{d}^{2}) - m_{s}^{2}f_{1}(m_{s}^{2}) 
\right \}
\left (
1+\frac{m_{j}^{2}}{2M_{W}^{2}}
\right )
\nonumber \\
& & \hskip-0.5cm
-\frac{m_{d}^{2}m_{s}^{2}}{2M_{W}^{2}}\left \{
f_{1}(m_{d}^{2}) - f_{1}(m_{s}^{2})
\right \}
+
\frac{(m_{d}^{2} + m_{s}^{2})m_{j}^{2}}{2M_{W}^{2}}\left \{
f_{2}(m_{d}^{2}) - f_{2}(m_{s}^{2})
\right \} \bigg ]\:,
\label{eq:c1mj}
\\
c_{2}(m_{j}) &=&
\frac{m_{d}m_{s}}{m_{d}^{2}-m_{s}^{2} } \bigg [
-  \left \{
 f_{1}(m_{d}^{2}) - f_{1}(m_{s}^{2}) 
\right \}
\left (
1+\frac{m_{j}^{2}}{2M_{W}^{2}}
\right )
\nonumber \\
& & -\frac{1}{2M_{W}^{2}}\left \{
m_{d}^{2} f_{1}(m_{d}^{2}) - m_{s}^{2} f_{1}(m_{s}^{2})
\right \}
+
\frac{m_{j}^{2}}{M_{W}^{2}}\left \{
f_{2}(m_{d}^{2}) - f_{2}(m_{s}^{2})
\right \} \bigg ]\:,
\label{eq:c2mj}
\\
c_{3}(m_{j}) &=&
\frac{m_{d}^{2}}{m_{d}^{2} - m_{s}^{2}} \bigg [
\left \{
f_{1}(m_{d}^{2}) - f_{1}(m_{s}^{2}) 
\right \}\left (
1+\frac{m_{j}^{2}}{2M_{W}^{2}} + \frac{m_{s}^{2}}{2M_{W}^{2}}
\right )
\nonumber \\
& & +\frac{m_{j}^{2}}{2M_{W}^{2}}\left \{
- \frac{m_{d}^{2}+m_{s}^{2}}{m_{d}^{2}}\:f_{2}(m_{d}^{2}) + 2f_{2}(m_{s}^{2})
\right \} \bigg ]\:,
\label{eq:c3mj}
\\
c_{4}(m_{j}) &=&
\frac{m_{s}^{2}}{m_{d}^{2} - m_{s}^{2}} \bigg [
\left \{
f_{1}(m_{d}^{2}) - f_{1}(m_{s}^{2}) 
\right \}\left (
1+\frac{m_{j}^{2}}{2M_{W}^{2}} + \frac{m_{d}^{2}}{2M_{W}^{2}}
\right )
\nonumber \\
& & +\frac{m_{j}^{2}}{2M_{W}^{2}}\left \{
 \frac{m_{d}^{2}+m_{s}^{2}}{m_{s}^{2}}\:f_{2}(m_{s}^{2}) - 2f_{2}(m_{d}^{2})
\right \} \bigg ]\:.
\label{eq:c4mj}
\end{eqnarray}

\subsection{Counter terms in the curved spacetime}
\label{subsec:countertermsinthecurvedspacetime}

So much for the counter terms in the flat Minkowski space and let us think about the 
generalization  to the curved background case.
The counter terms in the curved background come out  naturally by extending   
(\ref{eq:counterterm}) to a diffeomorphism invariant form, i.e., 
\begin{eqnarray}
\widehat{\cal L}_{\rm c.t.}
&=&
Z_{L} \: \overline{\Psi_{s}}_{L}\:  i  \gamma^{a}   \overleftrightarrow{\nabla}_{\mu}   \:  \Psi_{d\: L}\: {\tilde e}^{\mu}_{\:\:\: a}
+
Z_{R} \: \overline{\Psi_{s}}_{R}\:   i  \gamma^{a}  \overleftrightarrow{\nabla}_{\mu}   \: \Psi_{d\: R}\: {\tilde e}^{\mu}_{\:\:\: a}
\nonumber \\
& &
+ Z_{Y1}\: \sqrt{-e}\:\overline{\Psi_{s}}_{R}\:  m_{s} \Psi_{d\: L}
+
Z_{Y2}\: \sqrt{-e}\: \overline{\Psi_{s}}_{L}\: m_{d} \: \Psi_{d\: R}
\nonumber \\
& & +({\rm h.c.})\:,
\label{eq:countertermincurvedspace}
\end{eqnarray}
where the quark fields $\Psi _{d}$ and $\Psi_{s}$ are  weighted by $(-e)^{1/4}$ . 
The vierbein ${\tilde e^{\mu}}_{\:\:\:a} $ in Eq. (\ref{eq:countertermincurvedspace})
is expanded in $\kappa $ and thereby we get 
\begin{eqnarray}
\widehat{\cal L}_{\rm c.t.}
=\widehat{\cal L}_{\rm c.t.}^{(0)}+ \kappa \widehat{\cal L}_{\rm c.t.}^{(1)}+ \cdots \:, 
\end{eqnarray}
where $\widehat{\cal L}_{\rm c.t.}^{(0)}$ coincides with the flat space counter term (\ref{eq:counterterm}). 
The next term 
$\widehat{\cal L}_{\rm c.t.}^{(1)}$, on the other hand,  is  expressed as 
\begin{eqnarray}
\widehat{\cal L}_{\rm c.t.}^{(1)}
&=&
h^{\mu \nu} \bigg [
Z_{L} \: \overline{\Psi_{s}}_{L}\:  i  \gamma_{\: (\mu }   \overleftrightarrow{\nabla}_{\nu )}   \:  \Psi_{d\: L}\: 
+
Z_{R} \: \overline{\Psi_{s}}_{R}\:   i  \gamma_{\:(\mu}  \overleftrightarrow{\nabla}_{\nu )}   \: \Psi_{d\: R}\: 
\nonumber \\
& &
+ \frac{1}{
{\color{black}{D-2}}
} Z_{Y1}\: \eta_{\mu \nu} \:\overline{\Psi_{s}}_{R}\:  m_{s} \Psi_{d\: L}
+
\frac{1}{
{\color{black}{D-2}}
} Z_{Y2}\: \eta_{\mu \nu}\: \overline{\Psi_{s}}_{L}\: m_{d} \: \Psi_{d\: R}
\nonumber \\
& & +({\rm h.c.})\:\bigg ]\: ,
\label{eq:onegravitonvertexcounterterm}
\end{eqnarray}
and 
gives rise to the contribution depicted in Figure \ref{fig:counterterm} (b). 
As we will confirm later in Section \ref{sec:cancellation} explicitly, 
(\ref{eq:onegravitonvertexcounterterm})
eliminates the divergences 
in the one-graviton emission vertex-type diagrams (Figure \ref{fig:attachedtovertex} and 
Figure \ref{fig:attachedtopropagators}). 
It is 
to be noted 
that the renormalization constants, 
$Z_{L}$, $Z_{R}$, $Z_{Y1}$ and $Z_{Y2}$, 
are playing two roles: one is to render the self-energy type diagram 
(Figure \ref{fig:exercise2}) finite, and the other is to  
make the one-graviton emission vertex finite.  This is due to the fact that two counter term Lagrangians, 
(\ref{eq:counterterm}) and (\ref{eq:onegravitonvertexcounterterm}), should combine into the diffeomorphism invariant form (\ref{eq:countertermincurvedspace}).

It should be added herewith  that  Degrassi et al. \cite{degrasse} and Corian{\` o} et al.\cite{coriano1}   also 
previously discussed renormalization of the  vertex of
 (\ref{eq:dsgraviton}). They took a 
 sum of the 
vertex-type and self-energy type diagrams to find mutual  cancellation of divergences. 
 This cancellation is  consistent with our procedure 
of eliminating divergences simultaneously in both self-enegy type and vertex-type diagrams 
via $Z_{L}$, $Z_{R}$, $Z_{Y1}$ and $Z_{Y2}$\:.

Incidentally the coefficient $1/(D-2)$ in front of $Z_{Y1}$ and $Z_{Y2}$ in 
(\ref{eq:onegravitonvertexcounterterm}),  which comes from the formula 
(\ref{eq:rootminuse2}), gives rise to a finite deviation from $\displaystyle{\frac{1}{2}}Z_{Y1}$
and  $\displaystyle{\frac{1}{2}}Z_{Y2}$, namely, 
\begin{eqnarray}
\frac{1}{D-2}Z_{Y1}&=&\left \{ \frac{1}{2}-\frac{D-4}{2(D-2)} \right \} Z_{Y1}
\nonumber \\
&=&
\frac{1}{2}Z_{Y1}
{\color{black}{
+\frac{g^{2}}{(4\pi)^{2}}\sum_{j=t,c,u}(V_{\rm CKM})^{*}_{js}(V_{\rm CKM})_{jd}\frac{m_{j}^{2}}{M_{W}^{2}}\times 
\frac{1}{4}
}}\:, 
\label{eq:deviation1}
\end{eqnarray}
as we take the $D \to 4 $ limit. The same formula also applies to  $\displaystyle{\frac{1}{D-2}}Z_{Y2}$, 
i.e.,
\begin{eqnarray}
\frac{1}{D-2}Z_{Y2}
&=&
\frac{1}{2}Z_{Y2}
{\color{black}{
+\frac{g^{2}}{(4\pi)^{2}}\sum_{j=t,c,u}(V_{\rm CKM})^{*}_{js}(V_{\rm CKM})_{jd}\frac{m_{j}^{2}}{M_{W}^{2}}\times 
\frac{1}{4}
}}\: .
\label{eq:deviation2}
\end{eqnarray}

\section{Gravitational flavor-changing vertices}
\label{sec:gravitationalflavorchangingvertex}

Now that  we have the counter term Lagrangian (\ref{eq:onegravitonvertexcounterterm}) at our hand,  
we are  well-prepared to handle the divergences that appear in evaluating  (\ref{eq:dsgraviton}).  
The relevant Feynman diagrams for (\ref{eq:dsgraviton}) may be classified into two types: 
those with two internal propagators   (Figure  \ref{fig:attachedtovertex})
and those with three internal propagators (Figure \ref{fig:attachedtopropagators}) . 
The latter diagrams are expressed necessarily by double integrals with respect to  Feynman parameters, 
 while the 
former diagrams by  single ones.  We will keep the external quarks off-shell, refraining from 
using  the Dirac equation throughout.  We will never use any approximation as to the 
magnitude of the quark masses until Section \ref{sec:effectivelagrangian}, where 
the large top quark mass limit of the $d$-$s$-graviton vertex is investigated.

\subsection{A graviton attached to the charged current vertex}

\begin{figure}[b]
\begin{center}
\input{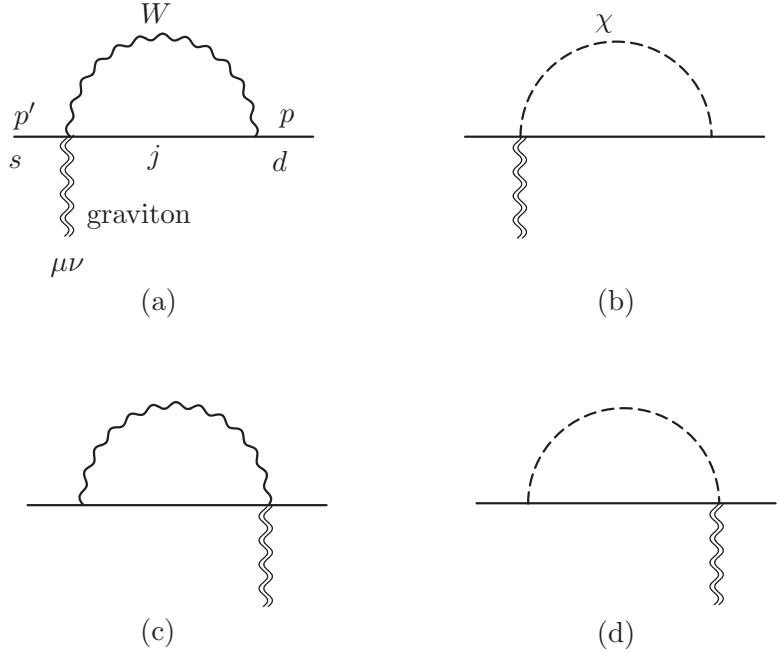}
\caption{Diagrams with a graviton attached to vertices}
\label{fig:attachedtovertex}
\end{center}
\end{figure}

Let us begin with the calculation of Figure  \ref{fig:attachedtovertex} in which graviton lines are 
attached to the charged current vertices.  Applications of the Feynman rules give us the following sum 
\begin{eqnarray}
\Gamma_{\mu \nu}^{({\rm Fig.}  \ref{fig:attachedtovertex})} (p, p^{\prime})
&=& 
\sum_{j=t,c,u} (V_{\rm CKM})_{js}^{*} (V_{\rm CKM})_{jd} \left \{ 
  {\cal G}^{(a)}_{\mu \nu} + {\cal G} ^{(b)}_{\mu \nu}   + {\cal G} ^{(c)}_{\mu \nu}
  +{\cal G}^{(d)}_{\mu \nu} 
 \right \}\:,
\end{eqnarray}
where for each diagram in Figure  \ref{fig:attachedtovertex}  we define respectively the integrations 
\begin{eqnarray}
{\cal G}_{\mu \nu}^{(a)}&\equiv&
\frac{i\kappa g^{2}}{4} \mu^{4-D} \int \frac{d^{D}q}{(2\pi)^{D}}\left (
\gamma_{\mu}\eta_{\nu \alpha} + \gamma_{\nu} \eta_{\mu \alpha}
\right )L \frac{i}{\gamma \cdot (p-q) -m_{j}}\gamma_{\beta} L
\frac{-i\; \eta^{\alpha \beta}}{q^{2} - M_{W}^{2}}\:,
\nonumber \\
\\
{\cal G}_{\mu \nu}^{(b)}&\equiv&
\frac{i\kappa g^{2}}{
{\color{black}{2(D-2)}}
}\cdot \frac{1}{M_{W}^{2}}
\mu ^{4-D}\:    \eta_{\mu \nu}  \int \frac{d^{D}q}{(2\pi)^{D}}
\frac{i}{q^{2} - M_{W}^{2}}
\nonumber \\
& & 
\hskip3cm 
\times (m_{j}R - m_{s}L) \frac{i}{\gamma \cdot (p-q) -m_{j}} (m_{j}L-m_{d}R)\:,
\label{eq:gmunu(b)}
\end{eqnarray}
\begin{eqnarray}
{\cal G}_{\mu \nu}^{(c)}&\equiv&
\frac{i\kappa g^{2}}{4} \mu^{4-D} \int \frac{d^{D}q}{(2\pi)^{D}}\: 
\gamma_{\beta}L \frac{i}{\gamma \cdot (p^{\: \prime}-q) -m_{j}}
\left (
\gamma_{\mu}\eta_{\nu \alpha} + \gamma_{\nu} \eta_{\mu \alpha}
\right )L
\frac{-i\; \eta^{\alpha \beta}}{q^{2} - M_{W}^{2}}\;,
\nonumber \\
\\
{\cal G}_{\mu \nu}^{(d)}&\equiv&
\frac{i\kappa g^{2}}{
{\color{black}{2(D-2)}}
}\cdot \frac{1}{M_{W}^{2}}
\mu ^{4-D}\:    \eta_{\mu \nu}  \int \frac{d^{D}q}{(2\pi)^{D}}
\frac{i}{q^{2} - M_{W}^{2}}
\nonumber \\
& & \hskip3cm
\times (m_{j}R - m_{s}L) \frac{i}{\gamma \cdot (p^{\prime} -q) -m_{j}} (m_{j}L-m_{d}R)\:.
\label{eq:gmunu(d)}
\end{eqnarray}
Note that the factor $1/(D-2)$ in front of  (\ref{eq:gmunu(b)}) and (\ref{eq:gmunu(d)}) is due to  the second 
term of (\ref{eq:rootminuse2}).
On comparing (\ref{eq:gmunu(b)}) and (\ref{eq:gmunu(d)}) with(\ref{eq:ibfirst}), 
one can immediately see a simple relation
\begin{eqnarray}
{\cal G}_{\mu \nu}^{(b)}=\frac{\kappa }{
{\color{black}{D-2}}
}\:\eta_{\mu \nu}\: {\cal S}^{(b)}(p), 
\hskip1cm
{\cal G}_{\mu \nu}^{(d)}=\frac{\kappa }{
{\color{black}{D-2}}
}\:\eta_{\mu \nu}\: {\cal S}^{(b)}(p^{\prime})\:.
\label{eq:relationtoselfenergy}
\end{eqnarray}
The evaluation of the above Feynman integrations is rather standard and we list up simply 
the results below:
\begin{eqnarray}
{\cal G}_{\mu \nu}^{(a)}
&=&
\frac{\kappa g^{2}}{(4\pi)^{2}}\: G_{1}(p^{2})\left \{
\gamma_{(\mu}p_{\nu)} -\frac{1}{2}\eta_{\mu \nu}\gamma \cdot p
\right \} L\:,
\label{eq:6.7}
\\
{\cal G}_{\mu \nu}^{(c)}
&=&
\frac{\kappa g^{2}}{(4\pi)^{2}}\: G_{1}(p^{\prime \: 2})\left \{
\gamma_{(\mu}{p_{\nu)}^{\prime}} -\frac{1}{2}\eta_{\mu \nu}\gamma \cdot p^{\prime}
\right \} L\:,
\label{eq:6.8}
\\
{\cal G}_{\mu \nu}^{(b)}
&=&
\frac{\kappa g^{2}}{(4\pi)^{2}} \cdot \frac{1}{4M_{W}^{2}}
\bigg [
\left \{
-G_{1}(p^{2}) + 
{\color{black}{\frac{1}{2}}}
\right\}
\eta_{\mu \nu} \gamma \cdot p \left (
m_{j}^{2}L+m_{s}m_{d}R \right )
\nonumber \\
& & \hskip3cm 
+
\left \{ 
G_{2}(p^{2})
{\color{black}{-1}}
\right \}
m_{j}^{2}\eta_{\mu \nu}\left (
m_{s}L+m_{d}R
\right ) \bigg ]   \:,
\label{eq:6.9}
\\
{\cal G}_{\mu \nu}^{(d)}
&=&
\frac{\kappa g^{2}}{(4\pi)^{2}} \cdot \frac{1}{4M_{W}^{2}}
\bigg [
\left \{ 
-G_{1}(p^{\prime \: 2})+
{\color{black}{\frac{1}{2}}}
\right \}
\eta_{\mu \nu} \gamma \cdot p^{\prime} \left (
m_{j}^{2}L+m_{s}m_{d}R \right )
\nonumber \\
& & \hskip3cm 
+
\left \{ 
G_{2}(p^{\prime \: 2})
{\color{black}{-1}}
\right \} 
m_{j}^{2}\eta_{\mu \nu}\left (
m_{s}L+m_{d}R
\right ) \bigg ]\:.
\label{eq:6.10}
\end{eqnarray}
The functions $G_{1}(p^{2})$ and $G_{2}(p^{2})$ are defined respectively by (\ref{eq:g1pp}) 
and (\ref{eq:g2pp}) in Appendix \ref{appendixb}. One can confirm that 
the formulae of ${\cal G}_{\mu \nu}^{(b)}$ and ${\cal G}_{\mu \nu}^{(d)}$ are nothing but 
those  obtained from (\ref{eq:sb})  by the relation (\ref{eq:relationtoselfenergy}).


\subsection{A graviton attached to the internal propagators}

\begin{figure}[hbt]
\begin{center}
\input{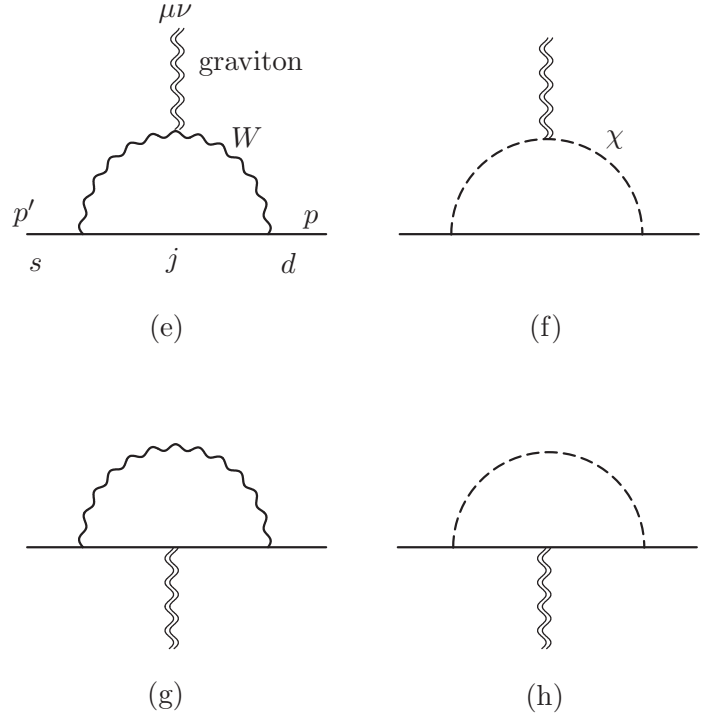}
\caption{Diagrams with a graviton attached to internal propagators}
\label{fig:attachedtopropagators}
\end{center}
\end{figure}

Another set of Feynman diagrams depicted in Figure \ref{fig:attachedtopropagators}
are those in which the graviton is attached to internal lines.  Let us define 
\begin{eqnarray}
\Gamma_{\mu \nu}^{({\rm Fig.}  \ref{fig:attachedtopropagators})} (p, p^{\prime})
&=& 
\sum_{j=t,c,u} (V_{\rm CKM})_{js}^{*} (V_{\rm CKM})_{jd} \left \{ 
  {\cal G}^{(e)}_{\mu \nu} + {\cal G} ^{(f)}_{\mu \nu}   + {\cal G} ^{(g)}_{\mu \nu}
  +{\cal G}^{(h)}_{\mu \nu} 
 \right \}\:,
\end{eqnarray}
where each term in the brackets on the right hand side corresponds to each diagram in 
 Figure \ref{fig:attachedtopropagators} and is given by 
\begin{eqnarray}
{\cal G}_{\mu \nu}^{(e)}
&\equiv& - \frac{\kappa g^{2}}{2} \mu^{4-D}\int \frac{d^{D}q}{(2\pi)^{D}}
\: \gamma^{\tau}L\frac{i}{\gamma \cdot q - m_{j}} \gamma^{\rho}L \cdot
\frac{-i}{(p-q)^{2} - M_{W}^{2}}\frac{-i}{(p^{\:\prime} -q)^{2} - M_{W}^{2}}
\nonumber \\
& & \times \bigg [ 
V_{\mu \nu  \sigma \tau \lambda \rho}\bigg \vert _{\xi =1}\:(p^{\prime}-q )^{\sigma} (p-q)^{\lambda}
+ 2M_{W}^{2} \: \eta_{\tau ( \mu} \: \eta_{\nu ) \rho}
\nonumber \\
& & 
-
2\: \eta_{\sigma ( \mu } \eta _{\nu ) \tau} \: \eta _{\lambda \rho}\: (p-q)^{\sigma} (p-q)^{\lambda} 
-
2\:\eta_{\sigma \tau}\eta_{\lambda ( \mu}\eta_{\nu ) \rho} (p^{\: \prime}-q)^{\lambda }(p^{\:\prime}
-q)^{\sigma} 
\bigg ]\:,
\\
{\cal G}_{\mu \nu}^{(f)}
&\equiv &
- \frac{\kappa g^{2}}{2}
\frac{1}{M_{W}^{2}}
\mu^{4-D} \int \frac{d^{D}q}{(2\pi)^{D}}(m_{j}R - m_{s}L)\frac{i}{\gamma \cdot q -m_{j}}
(m_{j}L - m_{d}R)
\nonumber \\
& & \times \frac{i}{(p^{\:\prime}-q)^{2} -M_{W}^{2}}\frac{i}{(p^{}-q)^{2} -M_{W}^{2}}
\nonumber \\
& & \times 
\left \{
(p^{\:\prime}-q)_{\mu}(p-q)_{\nu}+(p^{\:\prime} -q)_{\nu}(p-q)_{\mu}-
{\color{black}{\frac{2}{D-2}}}
\eta_{\mu \nu}M_{W}^{2}
\right \}\:, 
\end{eqnarray}
\begin{eqnarray}
{\cal G}_{\mu \nu}^{(g)}
&\equiv &
-\frac{\kappa g^{2}}{2}\mu^{4-D}\int \frac{d^{D}q}{(2\pi)^{D}}\frac{-i\eta_{\alpha \beta}}{q^{2}-M_{W}^{2}}
\nonumber \\
& & \times \gamma^{\alpha} L\frac{i}{\gamma \cdot (p^{\:\prime}-q)-m_{j}}
\nonumber \\
& & \times 
\left \{
\frac{1}{4}\gamma_{\mu}(p+p^{\:\prime}-2q)_{\nu} + 
\frac{1}{4}\gamma _{\nu}(p+p^{\:\prime}-2q)_{\mu} - 
{\color{black}{\frac{1}{D-2}}}
\eta_{\mu \nu}m_{j}
\right \}
\nonumber \\
& & \times \frac{i}{\gamma \cdot (p-q)-m_{j}}\gamma^{\beta} L\:,
\end{eqnarray}
\begin{eqnarray}
{\cal G}_{\mu \nu}^{(h)}
&\equiv &
-\frac{\kappa g^{2}}{2} \frac{1}{M_{W}^{2}} \mu^{4-D}\int \frac{d^{D}q}{(2\pi)^{D}}\frac{i}{q^{2}-M_{W}^{2}}
\nonumber \\
& & \times (m_{j}R -m_{s}L) \frac{i}{\gamma \cdot (p^{\:\prime}-q)-m_{j}}
\nonumber \\
& & \times 
\left \{
\frac{1}{4}\gamma_{\mu}(p+p^{\:\prime}-2q)_{\nu} + 
\frac{1}{4}\gamma _{\nu}(p+p^{\:\prime}-2q)_{\mu} - 
{\color{black}{\frac{1}{D-2}}}
\eta_{\mu \nu}m_{j}
\right \}
\nonumber \\
& & \times \frac{i}{\gamma \cdot (p-q)-m_{j}}(m_{j}L - m_{d}R)\:.
\end{eqnarray}

Now the calculations of the above integrals are again straightforward but tedious  
since  there are many types of gamma-matrix combinations and tensor structures. We 
 just list up our final formulas: 
\begin{eqnarray}
{\cal G}_{\mu \nu}^{(e)}
&=&
\frac{\kappa g^{2}}{(4\pi)^{2}} \bigg [ \: 
G_{3}(p, p^{\prime}) \eta_{\mu \nu} \gamma \cdot p + G_{3}(p^{\prime}, p)\eta _{\mu \nu} \gamma \cdot p^{\prime}
+ G_{4}(p, p^{\prime}) \gamma _{(\mu}p_{\nu)} +  G_{4}(p^{\prime}, p) \gamma _{(\mu}p^{\prime}_{\nu)}
\nonumber \\
& & + \left \{
-2f_{7}(p, p^{\prime}) p_{\mu}p_{\nu} + 2f_{8}(p, p^{\prime})p_{\mu}^{\prime}p_{\nu}^{\prime}
+2f_{9}(p, p^{\prime})p_{(\mu}p^{\prime}_{\nu )}
\right \} \gamma \cdot p
\nonumber \\
& & 
+ \left \{
-2f_{7}(p^{\prime}, p^{}) p^{\prime}_{\mu}p^{\prime}_{\nu} + 2f_{8}(p^{\prime}, p^{})p_{\mu}^{}p_{\nu}^{}
+2f_{9}(p^{\prime}, p^{})p_{(\mu}p^{\prime}_{\nu )}
\right \} \gamma \cdot p^{\prime}
\nonumber \\
& & + f_{10}(p, p^{\prime}) \gamma \cdot p^{\prime} \gamma _{(\mu}p_{\nu )} \gamma \cdot p
+ f_{10}(p^{\prime}, p) \gamma \cdot p^{\prime} \gamma _{(\mu}p^{\prime} _{\nu )} \gamma \cdot p
\: \bigg ] L\:, 
\end{eqnarray}
\begin{eqnarray}
{\cal G}_{\mu \nu}^{(f)}
&=&
\frac{\kappa g^{2}}{(4\pi)^{2}} \frac{1}{M_{W}^{2}}\bigg [ \: 
G_{5}(p, p^{\prime}) \eta_{\mu \nu} \gamma \cdot p + G_{5}(p^{\prime}, p)\eta _{\mu \nu} \gamma \cdot p^{\prime}
\nonumber \\
& & 
+ G_{6}(p, p^{\prime}) \gamma _{(\mu}p_{\nu)} +  G_{6}(p^{\prime}, p) \gamma _{(\mu}p^{\prime}_{\nu)}
\nonumber \\
& & + \left \{
-f_{11}(p^{\prime}, p) p_{\mu}^{\prime}p_{\nu}^{\prime} - f_{7}(p, p^{\prime})p_{\mu}^{}p_{\nu}^{}
+f_{12}(p^{\prime}, p)p_{(\mu}p^{\prime}_{\nu )}
\right \} \gamma \cdot p
\nonumber \\
& & 
+ \left \{
-f_{11}(p, p^{\prime}) p^{}_{\mu}p^{}_{\nu} -   f_{7}(p^{\prime}, p^{})p_{\mu}^{\prime}p_{\nu}^{\prime}
+ f_{12}(p, p^{\prime})p_{(\mu}p^{\prime}_{\nu )}
\right \} \gamma \cdot p^{\prime}
\: \bigg ] (m_{j}^{2}L+m_{s}m_{d}R)
\nonumber \\
& &
+ \frac{\kappa g^{2}}{(4\pi )^{2}} \frac{1}{M_{W}^{2}}\bigg [
G_{7}(p, p^{\prime})\eta_{\mu \nu}
\nonumber \\
& & 
+\left \{
f_{13}(p, p^{\prime})p_{\mu}^{\prime}p_{\nu}^{\prime}+ f_{13}(p^{\prime}, p)p^{}_{\mu}p^{}_{\nu}
- f_{14}(p, p^{\prime}) p_{(\mu}p_{\nu )}^{\prime}
\right \} \bigg ] m_{j}^{2}(m_{s}L+m_{d}R)\:, 
\label{eq:equationgmunuf}
\end{eqnarray}
\begin{eqnarray}
{\cal G}_{\mu \nu}^{(g)}
&=&
\frac{\kappa g^{2}}{(4\pi)^{2}} \bigg [ \: 
G_{8}(p, p^{\prime}) \eta_{\mu \nu} \gamma \cdot p + G_{8}(p^{\prime}, p)\eta _{\mu \nu} \gamma \cdot p^{\prime}
+ G_{9}(p, p^{\prime}) \gamma _{(\mu}p_{\nu)} +  G_{9}(p^{\prime}, p) \gamma _{(\mu}p^{\prime}_{\nu)}
\nonumber \\
& & + \left \{
2f_{17}(p, p^{\prime}) p_{\mu}p_{\nu} -2 f_{19}(p^{\prime}, p)p_{\mu}^{\prime}p_{\nu}^{\prime}
-2f_{21}(p, p^{\prime})p_{(\mu}p^{\prime}_{\nu )}
\right \} \gamma \cdot p
\nonumber \\
& & 
+ \left \{
2f_{17}(p^{\prime}, p^{}) p^{\prime}_{\mu}p^{\prime}_{\nu} -2 f_{19}(p, p^{\prime})p_{\mu}^{}p_{\nu}^{}
-2f_{21}(p^{\prime}, p^{})p_{(\mu}p^{\prime}_{\nu )}
\right \} \gamma \cdot p^{\prime}
\nonumber \\
& & + f_{22}(p, p^{\prime}) \gamma \cdot p^{\prime} \gamma _{(\mu}p_{\nu )} \gamma \cdot p
+ f_{22}(p^{\prime}, p) \gamma \cdot p^{\prime} \gamma _{(\mu}p^{\prime} _{\nu )} \gamma \cdot p
\: \bigg ] L\:,
\end{eqnarray}
\begin{eqnarray}
{\cal G}_{\mu \nu}^{(h)}
&=&
\frac{\kappa g^{2}}{(4\pi)^{2}} \frac{1}{M_{W}^{2}}\bigg [ \: 
G_{10}(p, p^{\prime}) \eta_{\mu \nu} \gamma \cdot p + G_{10}(p^{\prime}, p)\eta _{\mu \nu} \gamma \cdot p^{\prime}
\nonumber \\
& & 
+ G_{11}(p, p^{\prime}) \gamma _{(\mu}p_{\nu)} +  G_{11}(p^{\prime}, p) \gamma _{(\mu}p^{\prime}_{\nu)}
\nonumber \\
& & + \left \{
- \frac{1}{2}f_{17}(p, p^{\prime}) p_{\mu}p_{\nu}
 +\frac{1}{2} f_{28}(p^{\prime}, p)p_{\mu}^{\prime}p_{\nu}^{\prime}
- \frac{1}{2}f_{29}(p, p^{\prime})p_{(\mu}p^{\prime}_{\nu )}
\right \} \gamma \cdot p
\nonumber \\
& & 
+ \left \{
- \frac{1}{2}f_{17}(p^{\prime}, p) p^{\prime}_{\mu}p^{\prime}_{\nu} 
+ \frac{1}{2}  f_{28}(p, p^{\prime})p_{\mu}^{}p_{\nu}^{}
- \frac{1}{2}  f_{29}(p^{\prime}, p)p_{(\mu}p^{\prime}_{\nu )}
\right \} \gamma \cdot p^{\prime}
\nonumber \\
& & 
+\frac{1}{4}f_{22}(p, p^{\prime})\gamma \cdot p^{\prime} \gamma _{(\mu}p_{\nu )}\gamma \cdot p
+\frac{1}{4}f_{22}(p^{\prime}, p)\gamma \cdot p^{\prime} \gamma _{(\mu}p^{\prime} _{\nu )}\gamma \cdot p
\: \bigg ] (m_{j}^{2}L+m_{s}m_{d}R)
\nonumber \\
& &
+ \frac{\kappa g^{2}}{(4\pi )^{2}} \frac{1}{M_{W}^{2}}\bigg [
G_{12}(p, p^{\prime})\eta_{\mu \nu}
\nonumber \\
& & 
+\left \{
\frac{1}{2}f_{24}(p, p^{\prime})p_{\mu}p_{\nu}+
 \frac{1}{2}f_{24}(p^{\prime}, p)p^{\prime}_{\mu}p^{\prime}_{\nu}
+ \frac{1}{2} f_{23}(p, p^{\prime}) p_{(\mu}p_{\nu )}^{\prime}
\right \} 
\nonumber \\
& & 
+ \left \{
-\frac{1}{4}f_{20}(p, p^{\prime})\gamma_{(\mu}p_{\nu )}
-\frac{1}{4}f_{20}(p^{\prime}, p)\gamma_{(\mu}p^{\prime}_{\nu )}
\right \}\gamma \cdot p 
\nonumber\\
& &
+ \gamma \cdot p^{\prime} \left \{
-\frac{1}{4}f_{20}(p, p^{\prime})\gamma_{(\mu}p_{\nu )}
-\frac{1}{4}f_{20}(p^{\prime}, p)\gamma_{(\mu}p^{\prime}_{\nu )}
\right \}
\nonumber \\
& & 
+ \frac{1}{4}f_{25}(p, p^{\prime})\eta _{\mu \nu} \gamma \cdot p^{\prime} \gamma \cdot p
\bigg ] m_{j}^{2}(m_{s}L+m_{d}R)\:.
\label{eq:equationgmunuh}
\end{eqnarray}
Here we have introduced various kinds of Feynman parameters' integrations $f_{i}(p, p^{\prime})$, 
all of which are collected in Appendix \ref{sec:integralrepresentation1}. 
Some combinations $G_{i}(p, p^{\prime})$ $ (i=3, \cdots , 12)$ of $f_{i}(p, p^{\prime})$ are defined in Appendix \ref{appendixb}\:.

\section{Cancellation of ultraviolet divergences}
\label{sec:cancellation}

We are now ready  to sum up  the ultraviolet divergences that appear in the graviton emission 
vertex
\begin{eqnarray}
\Gamma _{\mu \nu}(p, p^{\prime}) 
\equiv 
\Gamma_{\mu \nu}^{({\rm Fig.}  \ref{fig:attachedtovertex})} (p, p^{\prime}) 
+
\Gamma_{\mu \nu}^{({\rm Fig.}  \ref{fig:attachedtopropagators})}  (p, p^{\prime})   \:.
\end{eqnarray}
As we see in the formulae of Appendix \ref{appendixb},  the quantities $G_{1}(p^{2})$, 
$G_{2}(p^{2})$ and $G_{i}(p, p^{\prime})$  $(i=3, \cdots , 11)$ all have a pole term $1/(D-4)$. 
In (\ref{eq:6.7}), 
  for example,  we notice that $G_{1}(p^{2})$ is not 
accompanied by $m_{j}^{2}$ or any $j$-dependent factors 
and therefore the pole term in $G_{1}(p^{2})$ in  (\ref{eq:6.7}) 
do not survive the $j$-($=t$, $c$ and  $u$ ) summation because of the unitarity 
relation (\ref{eq:ckmunitarity}). The same  comment applies to many of  the other pole terms. 
Namely, the pole terms survive the $j$-summation only when multiplied  by $j$-dependent 
factors such as $m_{j}$. It is noteworthy 
 that not only the divergences in Figures  \ref{fig:attachedtovertex} (a) and \ref{fig:attachedtovertex} (c) 
 but also   those of Figures  \ref{fig:attachedtopropagators} (e) and  \ref{fig:attachedtopropagators}  (g) 
  disappear after  the summation over $j$. 
Putting  remaining ultraviolet divergent terms all together, we end up with the following expression 
for the divergences, 
\begin{eqnarray}
& & \hskip-1cm
\Gamma _{\mu \nu}(p, p^{\prime}) 
\nonumber \\
&=&  \frac{\kappa g^{2}}{(4\pi)^{2}}  \sum _{j}(V_{\rm CKM})_{js}^{*}(V_{\rm CKM})_{jd}
\bigg [
\nonumber \\
& & + \left (
-\frac{1}{4}\cdot \frac{1}{D-4}\right )
\gamma_{(\mu}(p+p^{\prime})_{\nu )} 
\:\frac{m_{j}^{2}}{M_{W}^{2}}\:L
+ \eta_{\mu \nu}\: 
\left (
\frac{1}{2}\cdot \frac{1}{D-4}
\right )
\: \frac{m_{j}^{2}}{M_{W}^{2}}
\left ( m_{s}L + m_{d}R \right )
\nonumber \\
& & +({\rm finite \:\:\: terms}) \bigg ]\:.
\label{eq:uvdivergencesinthevertex}
\end{eqnarray}
Note that divergences proportional to $\eta_{\mu \nu}\: \gamma \cdot ( p + p^{\prime})$ 
disappear in (\ref{eq:uvdivergencesinthevertex}) via mutual cancellation.

The divergences in (\ref{eq:uvdivergencesinthevertex}) should be compared with 
 the counter term contributions 
$\Gamma^{\rm c.t.}_{\mu \nu}(p, p^{\prime})$ due to  (\ref{eq:onegravitonvertexcounterterm})
(Figure \ref{fig:counterterm} (b)),    namely, 
\begin{eqnarray}
& &\hskip-1cm
\Gamma^{\rm c.t.}_{\mu \nu}(p, p^{\prime})
\nonumber \\
&=& 
\frac{\kappa }{2}Z_{L} \gamma_{\:(\mu}(p+p^{\prime})_{\: \nu)}\: L
+
\frac{\kappa }{2}Z_{R} \gamma_{\:(\mu}(p+p^{\prime})_{\: \nu)}\: R
\nonumber \\
& & 
+
\frac{\kappa }{
{\color{black}{D-2}}
}Z_{Y1}\eta _{\mu \nu} m_{s}L 
+
 \frac{\kappa }{
 {\color{black}{D-2}}
 }Z_{Y2}\eta _{\mu \nu} m_{d}R
\label{eq: vertexcounterterm}
\\
& =&      \frac{\kappa g^{2}}{(4\pi )^{2}} 
\sum _{j}(V_{\rm CKM})_{js}^{*}(V_{\rm CKM})_{jd}
\nonumber \\
& & \hskip0.4cm \times 
\bigg [ \frac{1}{4}\cdot \frac{1}{D-4} \cdot \frac{m_{j}^{2}}{M_{W}^{2}}\:\gamma_{\: (\mu}(p+p^{\prime})_{\nu )}L
- \frac{1}{2}\cdot \frac{1}{D-4} \cdot \frac{m_{j}^{2}}{M_{W}^{2}}\eta_{\mu \nu} (m_{s}L +m_{d}R ) 
\bigg ]
\nonumber \\
& & +({\rm finite \:\:\: terms})\:.
\label{eq:countertermcontributions}
\end{eqnarray}
Apparently, the $D=4$ pole terms in (\ref{eq:uvdivergencesinthevertex}) are cancelled out 
by the corresponding counter term contributions in (\ref{eq:countertermcontributions}).
This type of cancellation is the same as what has been known for long time 
in the $d$-$s$-$\gamma$ vertex  analyses \cite{inamilim, botella}\:.

We have thus confirmed the finiteness of the sum 
\begin{eqnarray}
\Gamma_{\mu \nu}^{\rm ren}(p, p^{\prime})
=
 \Gamma_{\mu \nu} (p, p^{\prime}) + \Gamma _{\mu \nu}^{\rm {c.t.}}(p, p^{\prime}),
\label{eq:renormaliseddsgravitonvertex}
 \end{eqnarray} 
 which we now call renormalized $d$-$s$-graviton vertex.
  The S-matrix element for the process (\ref{eq:dsgraviton}) is now given a finite value 
 through (\ref{eq:renormaliseddsgravitonvertex}). 
 When we deal with  S-matrix elements in general, renormalization of external lines has usually 
 to be    taken into account. In our case, however, the renormalized two-point function  
 $\Sigma _{\rm ren}(p)$ vanishes due to the subtraction conditions (\ref{eq:subtractioncondition})  
once we put external $d$- and $s$-quarks on the mass shell, and therefore 
it does not seem to affect the   S-matrix element of (\ref{eq:dsgraviton}). 
This, however, does not necessarily mean that 
the external line renormalization 
is not playing any role in the computation of the S-matrix. 
Actually  recall that the renormalization constants 
$Z_{L}$, $Z_{R}$, $Z_{Y1}$ and $Z_{Y2}$ contain finite terms $c_{1}(m_{j})$, $c_{2}(m_{j})$, 
$c_{3}(m_{j})$, and $c_{4}(m_{j})$, respectively, as we see in Eqs. (\ref{eq:zl}) through 
(\ref{eq:zy2}).  These  finite  terms are taken over in 
$\Gamma_{\mu \nu}^{\rm ren}(p, p^{\prime})$ after  the pole term cancellation in 
 (\ref{eq:renormaliseddsgravitonvertex}). 
Also remember  that these terms are all shared by the two-point function 
$\Sigma_{\rm ren}(p)$  as we see in   (\ref{eq:renormaliseddstransitionamplitudes}).  
The finite terms $c_{i}(m_{j})\:\:\:(i=1, \cdots , 4)$ in $\Gamma_{\mu \nu}^{\rm ren}(p, p^{\prime})$ and   
those in $\Sigma_{\rm ren}(p)$ are the two sides of the same coin and are 
closely   linked.   In this sense the  two-point function  $\Sigma_{\rm ren}(p)$ 
is an integral part  in computing  S-matrix elements. 

\section{Ward-Takahashi identity }
\label{sec:wardtakahashi}

In the present paper the gravitational field is always  treated as an external field and the 
invariance properties associated with the general coordinate transformation are reflected 
in the Feynman integrals.   Such invariance properties ought to  be  expressed    in    the 
form of Ward-Takahashi identities among Green's functions, whose field theoretical 
derivation, however,  would be rather 
involved due to the existence of  unphysical modes. Here we would like to 
use  a much more  naive  ``bottom-up" method. Namely,  we deal with the linear combinations
\begin{eqnarray}
{\cal G}_{\mu \nu}^{(X)}- \frac{1}{2}\eta_{\mu\nu} \eta^{\lambda \rho} {\cal G}_{\lambda \rho}^{(X)}, 
\hskip1cm (X=a, b, \cdots , h)
\label{eq:wedealwith}
\end{eqnarray}
correspondingly to (\ref{eq:conservedenergymomentumtensor}), 
multiply the Feynman integrals (\ref{eq:wedealwith})
  by $(p-p^{\prime})^{\mu}$,   shuffle the integrands in an algebraic way without performing 
  the integrations and eventually associate   (\ref{eq:wedealwith}) with  the  integrals of $d \to s $ self-energy type   diagrams, ${\cal S}^{(a)}$ of (\ref{eq:ia1first})  and ${\cal S}^{(b)}$ of (\ref{eq:ibfirst}).
 The identity we thus found after all is 
 \begin{eqnarray}
 & & \hskip-1cm 
 (p-p^{\prime})^{\mu}\left \{
 \Gamma_{\mu \nu} (p, p^{\prime}) -\frac{1}{2}\eta_{\mu \nu}\eta^{\lambda \rho}
 \Gamma_{\lambda \rho} (p, p^{\prime})
 \right \}
 \nonumber \\
 &=&\kappa \bigg \{ p^{\prime}_{\nu}\: \Sigma(p) - p_{\nu}\: \Sigma(p^{\prime})
 +
 \frac{1}{4}\Sigma(p^{\prime}) \gamma \cdot (p-p^{\prime})\gamma_{\nu}
 +
 \frac{1}{4} \gamma_{\nu}\gamma \cdot (p-p^{\prime})\Sigma(p)
 \bigg \}\:.
 \label{eq:wtidentitybare}
 \end{eqnarray}

Very   curiously, the counter terms (\ref{eq:selfenergycounterterm}) and 
  (\ref{eq: vertexcounterterm}) also satisfy the   identity of the same form, namely, 
 \begin{eqnarray}
 & & 
 (p-p^{\prime})^{\mu}\left \{
 \Gamma^{\rm c.t.}_{\mu \nu}(p, p^{\prime})  -\frac{1}{2}\eta_{\mu \nu}\eta^{\lambda \rho}
 \Gamma^{\rm c.t.}_{\lambda \rho} (p, p^{\prime})
 \right \}
 \nonumber \\
 &=&\kappa \bigg \{ p^{\prime}_{\nu}\: \Sigma_{\rm c.t.}(p) - p_{\nu}\: \Sigma_{\rm c.t.}(p^{\prime})
 +
 \frac{1}{4}\Sigma_{\rm c.t.}(p^{\prime}) \gamma \cdot (p-p^{\prime})\gamma_{\nu}
 +
 \frac{1}{4} \gamma_{\nu}\gamma \cdot (p-p^{\prime})\Sigma_{\rm c.t.}(p)
 \bigg \}\:.
 \nonumber \\
 \label{eq:wtidentitycounterterm}
 \end{eqnarray}
Combining   (\ref{eq:wtidentitybare})  and   (\ref{eq:wtidentitycounterterm})
we find that the renormalized quantities  
(\ref{eq:renormalizedselfenergydefined})  and  (\ref{eq:renormaliseddsgravitonvertex}) 
also satisfy the same identity, 
 \begin{eqnarray}
 & & 
 (p-p^{\prime})^{\mu}\left \{
 \Gamma^{\rm ren}_{\mu \nu}(p, p^{\prime})  -\frac{1}{2}\eta_{\mu \nu}\eta^{\lambda \rho}
 \Gamma^{\rm ren}_{\lambda \rho} (p, p^{\prime})
 \right \}
 \nonumber \\
 &=&\kappa \bigg \{ p^{\prime}_{\nu}\: \Sigma_{\rm ren}(p) - p_{\nu}\: \Sigma_{\rm ren}(p^{\prime})
 +
 \frac{1}{4}\Sigma_{\rm ren}(p^{\prime}) \gamma \cdot (p-p^{\prime})\gamma_{\nu}
 +
 \frac{1}{4} \gamma_{\nu}\gamma \cdot (p-p^{\prime})\Sigma_{\rm ren}(p)
 \bigg \}\:.
 \nonumber \\
 \label{eq:wtidentityrenormalised}
 \end{eqnarray}
We have checked the consistency of our Feynman integrations by referring to  these identities.
Note that, if external quarks are on the mass-shell, the identity 
(\ref{eq:wtidentityrenormalised}) reduces to the transversality condition
 \begin{eqnarray}
 & & \hskip-1cm 
 (p-p^{\prime})^{\mu}\left \{
 \Gamma^{\rm ren}_{\mu \nu} (p, p^{\prime}) -\frac{1}{2}\eta_{\mu \nu}\eta^{\lambda \rho}
 \Gamma^{\rm ren}_{\lambda \rho} (p, p^{\prime})
 \right \}
 =0, \hskip2cm ({\rm on\:\:shell})\:,
 \label{eq:transversalityequation}
  \end{eqnarray}
due to the  subtraction conditions
(\ref{eq:subtractioncondition})\:.

In the present paper all of the Feynman integrations are  performed in the 't Hooft-Feynman gauge. 
For the above-mentioned analyses of the Ward-Takahashi identity, however,  we have confirmed 
explicitly that 
Eqs. (\ref{eq:wtidentitybare}) through (\ref{eq:transversalityequation}) are all valid 
in the general $R_{\xi}$ gauge.
Incidentally the Ward-Takahashi identity associated with (\ref{eq:dsgraviton}) was also worked out  by
Corian{\` o} et al.\cite{coriano1}    Our identity (\ref{eq:wtidentityrenormalised}) is essentially 
the same as  theirs except for  the difference due to the weight factor $(-e)^{1/4}$ on the 
quark fields.


\section{The  large top quark mass limit}
\label{sec:effectivelagrangian}

Looking at the results of the graph calculations in Section \ref{sec:gravitationalflavorchangingvertex}, 
we notice immediately that the squared masses  of the intermediate  quarks,  
i.e., $m_{j}^{2}, \:\:(j=u, c, t)$ appear  explicitly in (\ref{eq:6.9}), (\ref{eq:6.10}), 
(\ref{eq:equationgmunuf}) and (\ref{eq:equationgmunuh})  besides those in the Feynman integrations. 
The origin of this $m_{j}$-dependence is traced back to the coupling of the unphysical scalar field 
to the quarks. Furthermore we notice  that the renormalization constants (\ref{eq:zl}),  (\ref{eq:zy1}) 
and (\ref{eq:zy2}), have the factor $m_{j}^{2}/M_{W}^{2}$ 
as coefficients of the $D=4$ pole terms.   The finite terms $c_{i}(m_{j}) \:\:\:(i=1, \cdots , 4)$ 
in the renormalization constants 
also contain $m_{j}^{2}/M_{W}^{2}$ explicitly, as we see in (\ref{eq:c1mj}), (\ref{eq:c2mj}),
 (\ref{eq:c3mj}) and  (\ref{eq:c4mj}). We are very much interested in whether or not such an  explicit 
 linear dependence on 
 $m_{j}^{2}/M_{W}^{2}$ could survive the summation of all the diagrams, for the large factor  
  $m_{t}^{2}/M_{W}^{2} \:\: (\approx 4.62) $ of the top quark's 
 would have  an enhancement effect  on the process (\ref{eq:dsgraviton})\:.

Up to Section \ref{sec:wardtakahashi}, we have never used any approximation with respect to
the magnitude of the quark masses. 
In the present Section, however, since  we are going to pay attention to 
  the large top quark mass behavior of our loop calculations, we suppose 
 that we can neglect    all the other quark masses    together with  external momenta squared, 
$p^{2}$,   $p^{\prime 2}$ and $(p-p^{\prime})^{2}$.  
We now have to perform the Feynman parameters' integrations explicitly   
under this approximation, which  
can be done in a straightforward way.  After  such calculations, however, our formulas would be   
extremely  cluttered and it is easy for us to lose sight of the essential points.  
Therefore in order to have a clear  insight into  our calculation, 
we  suppose an additional  relation 
$m_{t}^{2} \gg M_{W}^{2}$ . 
This relation  is used only to inspect  the structure of power series expansion with respect to 
$m_{t}^{2}/M_{W}^{2}$   .

As mentioned above, the most dominant terms in the large top quark mass limit come from the 
unphysical scalar exchange diagrams, i.e., Figures \ref{fig:attachedtovertex} (b), 
\ref{fig:attachedtovertex} (d),   \ref{fig:attachedtopropagators} (f) and 
\ref{fig:attachedtopropagators} (h).   
Therefore we collect all those
 terms that contain $m_{t}^{2}$ in front,   take the $m_{t}^{2}/M_{W}^{2} \to \infty $ limit 
 in the parameter integration, and then arrive at the following formula
\begin{eqnarray}
\Gamma_{\mu \nu} (p, p^{\prime})
& = &  \:
 \frac{\kappa g^{2}}{(4\pi)^{2}}\: \frac{m_{t}^{2}}{M_{W}^{2}}\:  (V_{\rm CKM})^{*}_{ts}(V_{\rm CKM})_{td}
\bigg [ 
\nonumber \\
& &
+ \left \{
-\frac{1}{4}\cdot \frac{1}{D-4} -\frac{1}{8}\:{\rm log}\left (
\frac{m_{t}^{2}}{4\pi \mu^{2}e^{-\gamma_{E}}} \right )+ \frac{3}{16}
\right \} \gamma_{(\mu}(p+p^{\prime})_{\nu )}L
\nonumber \\
& &  + \left \{
\frac{1}{2}\cdot \frac{1}{D-4} + \frac{1}{4}\:{\rm log}\left (
\frac{m_{t}^{2}}{4\pi \mu^{2}e^{-\gamma_{E}}} \right )- \frac{1}{2}
\right \} \eta_{\mu \nu}(m_{s}L +m_{d}R)
\nonumber \\
& & 
+ {\cal O}\left ( \frac{1}{m_{t}^{2}}  \right )  \bigg ]\;. 
\label{eq:largetopquarkmasslimitgammamunu}
\end{eqnarray}
Note that terms proportional to $\eta_{\mu \nu}\gamma \cdot (p+p^{\prime})$ have disappeared  
in (\ref{eq:largetopquarkmasslimitgammamunu}) after mutual cancellation.

The pole terms at $D=4$ 
in (\ref{eq:largetopquarkmasslimitgammamunu})
are to be cancelled  by the corresponding ones  in the counter terms
that  also contain 
$m_{t}^{2}/M_{W}^{2}$  in front.   
The renormalization constants may be expressed     in the following way: 
\begin{eqnarray}
Z_{L}
&\approx &
\frac{g^{2}}{(4\pi)^{2}}
(V_{\rm CKM})^{*}_{ts}(V_{\rm CKM})_{td}
\:\:  \frac{m_{t}^{2}}{M_{W}^{2}} \left (
\frac{1}{2} \cdot  \frac{1}{D-4} \:
- \tilde c_{1} 
\right )\;,
\label{eq:approximatezl}
\\
Z_{R}
& \approx &
\frac{g^{2}}{(4\pi)^{2}}
(V_{\rm CKM})^{*}_{ts}(V_{\rm CKM})_{td}
\:\:  \frac{m_{t}^{2}}{M_{W}^{2}} \times (- \tilde c_{2})    \;,
\label{eq:approximatezr}
\end{eqnarray}
\begin{eqnarray}
Z_{Y1}
& \approx &
\frac{g^{2}}{(4\pi)^{2}}
(V_{\rm CKM})^{*}_{ts}(V_{\rm CKM})_{td}
\:\:  \frac{m_{t}^{2}}{M_{W}^{2}}\left (
-    \frac{1}{D-4} \:
- \tilde c_{3} 
\right )\;,
\label{eq:approximatezy1}
\\
Z_{Y2}
&\approx &
\frac{g^{2}}{(4\pi)^{2}}
(V_{\rm CKM})^{*}_{ts}(V_{\rm CKM})_{td}
\:\:  \frac{m_{t}^{2}}{M_{W}^{2}} \left (
-   \frac{1}{D-4} \: 
- \tilde c_{4} 
\right )  \;.
\label{eq:approximatezy2}
\end{eqnarray}
Here  four quantities $\tilde c_{i}\:\:\:(i=1, \cdots , 4)$ are extracted respectively 
from $c_{i}(m_{t})\:\:\:
( i=1, \cdots , 4)$ as coefficients of those proportional to $m_{t}^{2}/M_{W}^{2}$, namely, 
\begin{eqnarray}
\tilde c_{1}&=&
\frac{1}{m_{d}^{2}-m_{s}^{2} } \bigg [
- \frac{1}{2}\left \{
m_{d}^{2} f_{1}(m_{d}^{2}) - m_{s}^{2}f_{1}(m_{s}^{2}) 
\right \}
\label{eq:c1tilde}
+
\frac{(m_{d}^{2} + m_{s}^{2})}{2}\left \{
f_{2}(m_{d}^{2}) - f_{2}(m_{s}^{2})
\right \} \bigg ]\:,
\nonumber
\\
\\
\tilde c_{2}&=&
\frac{m_{d}m_{s}}{m_{d}^{2}-m_{s}^{2} } \bigg [
-  \frac{1}{2}\left \{
 f_{1}(m_{d}^{2}) - f_{1}(m_{s}^{2}) 
\right \}
\label{eq:c2tilde}
+
\left \{
f_{2}(m_{d}^{2}) - f_{2}(m_{s}^{2})
\right \} \bigg ]\:,
\\
\tilde c_{3}&=&
\frac{m_{d}^{2}}{m_{d}^{2} - m_{s}^{2}} \bigg [
\frac{1}{2} \left \{
f_{1}(m_{d}^{2}) - f_{1}(m_{s}^{2}) 
\right \}
+\frac{1}{2}\left \{
- \frac{m_{d}^{2}+m_{s}^{2}}{m_{d}^{2}}\:f_{2}(m_{d}^{2}) + 2f_{2}(m_{s}^{2})
\right \} \bigg ]\:,
\label{eq:c3tilde}
\\
\tilde c_{4}&=&
\frac{m_{s}^{2}}{m_{d}^{2} - m_{s}^{2}} \bigg [
\frac{1}{2}   \left \{
f_{1}(m_{d}^{2}) - f_{1}(m_{s}^{2}) 
\right \}
+\frac{1}{2}\left \{
 \frac{m_{d}^{2}+m_{s}^{2}}{m_{s}^{2}}\:f_{2}(m_{s}^{2}) - 2f_{2}(m_{d}^{2})
\right \} \bigg ]\:.
\label{eq:c4tilde}
\end{eqnarray}
Recall that the original definitions of $f_{1}(p^{2})$ and $f_{2}(p^{2})$ contain $m_{j}$ as we see in 
(\ref{eq:f1(pp)}) and (\ref{eq:f2(pp)}). 
Here, however, we understand that   
 all $m_{j}$'s in $f_{1}$ and $f_{2}$ in (\ref{eq:c1tilde}),   (\ref{eq:c2tilde}),   (\ref{eq:c3tilde}) and 
(\ref{eq:c4tilde})   have been replaced by the top quark mass  $m_{t}$, 
namely, 
\begin{eqnarray}
f_{1}( p^{2})
&=&\int _{0}^{1} dx \: 
(1-x) {\rm log} \left \{
\frac{-x(1-x)p^{2}+xm_{t}^{2}+(1-x)M_{W}^{2}}{4\pi \mu^{2}e^{-\gamma_{E}}}
\right \}\;, 
\label{eq:f1(ppmt)}
\\
f_{2}( p^{2})
&=&\int _{0}^{1} dx \: 
 {\rm log} \left \{
\frac{-x(1-x)p^{2}+xm_{t}^{2}+(1-x)M_{W}^{2}}{4\pi \mu^{2}e^{-\gamma_{E}}}
\right \}\;.
\label{eq:f2(ppmt)}
\end{eqnarray}
The approximate equality $`` \approx "$ in Eqs. (\ref{eq:approximatezl}), (\ref{eq:approximatezr}), 
(\ref{eq:approximatezy1})  and (\ref{eq:approximatezy2}) means that we have simply collected 
terms containing  
$m_{t}^{2}/M_{W}^{2}$ as an overall factor    without going into the details of the 
$m_{t}$-dependence of $\tilde c_{i} \:\:\;(i=1, \cdots , 4)$   through $f_{1}$ and $f_{2}$.

We now look at the four quantities $\tilde c_{i}\:\: (i=1, \cdots , 4)$ more closely,  namely, 
their $m_{t}$-dependence entering through $f_{1}$  and $f_{2}$.
Taking the limit $m_{t} \to \infty $ 
while neglecting $M_{W}^{2}$ and $p^{2}$
in  (\ref{eq:f1(ppmt)}) and   (\ref{eq:f2(ppmt)}),   
we find immediately the following asymptotic behavior 
 \begin{eqnarray}
& &  f_{1}(p^{2}) = -\frac{3}{4}
  +\frac{1}{2}{\rm log}\left ( \frac{m_{t}^{2}}{4\pi \mu ^{2}e^{-\gamma_{E}}} \right )
  + {\cal O}\left ( \frac{M_{W}^{2}}{m_{t}^{2}},\:  \frac{p^{2}}{m_{t}^{2}} \right )\:, 
  \label{eq:f1asymptoticformula}
 \\
& &  f_{2}(p^{2}) = -1
  + {\rm log}\left ( \frac{m_{t}^{2}}{4\pi \mu ^{2}e^{-\gamma_{E}}} \right )
    + {\cal O}\left ( \frac{M_{W}^{2}}{m_{t}^{2}},\:  \frac{p^{2}}{m_{t}^{2}} \right )  \:.
    \label{eq:f2asymptoticformula}
 \end{eqnarray}
 Inserting     (\ref{eq:f1asymptoticformula})  and   (\ref{eq:f2asymptoticformula}) into Eqs. 
(\ref{eq:c1tilde}), (\ref{eq:c2tilde}), (\ref{eq:c3tilde})   and    (\ref{eq:c4tilde}), we obtain  the large-$m_{t}$ 
behavior of the four quantities $\tilde c_{i}\:\:\: (i=1, \cdots , 4)$ as follows
\begin{eqnarray}
& & {\tilde c}_{1}
=
\frac{3}{8} -\frac{1}{4} {\rm log}\left ( \frac{m_{t}^{2}}{4\pi \mu ^{2}e^{-\gamma_{E}}} 
\right ) + {\cal O}\left ( \frac{1}{m_{t}^{2}} \right )\:, 
\\
& & {\tilde c}_{2}= {\cal O}\left ( \frac{1}{m_{t}^{2}} \right )\:, 
\\
& & {\tilde c}_{3}
=
- \frac{1}{2} + \frac{1}{2} {\rm log}\left ( \frac{m_{t}^{2}}{4\pi \mu ^{2}e^{-\gamma_{E}}} 
\right ) + {\cal O}\left ( \frac{1}{m_{t}^{2}} \right )\:, 
\\
& & {\tilde c}_{4}
=
 - \frac{1}{2} + \frac{1}{2} {\rm log}\left ( \frac{m_{t}^{2}}{4\pi \mu ^{2}e^{-\gamma_{E}}} 
\right ) + {\cal O}\left ( \frac{1}{m_{t}^{2}} \right )\:.
\end{eqnarray}
By putting   these formulas into  (\ref{eq:approximatezl}), (\ref{eq:approximatezr}), 
(\ref{eq:approximatezy1})   and  (\ref{eq:approximatezy2}), the four renormalization constants turn out 
in the leading order in $m_{t}^{2}/M_{W}^{2}$   to be 
\begin{eqnarray}
Z_{L}
& =  &
\frac{g^{2}}{(4\pi)^{2}}
(V_{\rm CKM})^{*}_{ts}(V_{\rm CKM})_{td}
\:\:  \frac{m_{t}^{2}}{M_{W}^{2}}
\nonumber \\
& & \hskip1cm \times  \bigg \{
\frac{1}{2} \cdot  \frac{1}{D-4} \:
- \frac{3}{8}  + \frac{1}{4} {\rm log}\left ( \frac{m_{t}^{2}}{4\pi \mu ^{2}e^{-\gamma_{E}}} 
\right )
+ {\cal O}\left ( \frac{1}{m_{t}^{2}} \right )
\bigg \}\;,
\label{eq:approximatezlXXX}
\\
Z_{R}
& = &
\frac{g^{2}}{(4\pi)^{2}}
(V_{\rm CKM})^{*}_{ts}(V_{\rm CKM})_{td}
\:\:  \frac{m_{t}^{2}}{M_{W}^{2}} \times    {\cal O}\left ( \frac{1}{m_{t}^{2}} \right )      \;,
\label{eq:approximatezrXXX}
\\
Z_{Y1}
& = &
\frac{g^{2}}{(4\pi)^{2}}
(V_{\rm CKM})^{*}_{ts}(V_{\rm CKM})_{td}
\:\:  \frac{m_{t}^{2}}{M_{W}^{2}}
\nonumber \\
& & \hskip1cm \times  \left \{
-    \frac{1}{D-4} 
+  \frac{1}{2} - \frac{1}{2} {\rm log}\left ( \frac{m_{t}^{2}}{4\pi \mu ^{2}e^{-\gamma_{E}}} 
\right ) + {\cal O}\left ( \frac{1}{m_{t}^{2}} \right )
\right \} \;,
\label{eq:approximatezy1XXX}
\\
Z_{Y2}
&  =  &
\frac{g^{2}}{(4\pi)^{2}}
(V_{\rm CKM})^{*}_{ts}(V_{\rm CKM})_{td}
\:\:  \frac{m_{t}^{2}}{M_{W}^{2}} 
\nonumber \\
& & \hskip1cm \times \left \{
-   \frac{1}{D-4} 
 +  \frac{1}{2} -  \frac{1}{2} {\rm log}\left ( \frac{m_{t}^{2}}{4\pi \mu ^{2}e^{-\gamma_{E}}} 
\right ) + {\cal O}\left ( \frac{1}{m_{t}^{2}} \right )
\right \}   \;.
\label{eq:approximatezy2XXX}
\end{eqnarray}
We  thus  find that the  counterterm  contribution to the vertex (\ref{eq: vertexcounterterm}) is given 
in the $m_{t} \to \infty $ limit by 
 \begin{eqnarray}
\Gamma^{\rm c.t.}_{\mu \nu}(p, p^{\prime})
&=& 
\frac{\kappa }{2}Z_{L} \gamma_{\:(\mu}(p+p^{\prime})_{\: \nu)}\: L
+
\frac{\kappa }{2}Z_{R} \gamma_{\:(\mu}(p+p^{\prime})_{\: \nu)}\: R
\nonumber \\
& & 
+
\frac{\kappa }{D-2}
Z_{Y1}\eta _{\mu \nu} m_{s}L 
+
 \frac{\kappa }{D-2}
 Z_{Y2}\eta _{\mu \nu} m_{d}R
\nonumber \\
&=&
\frac{\kappa g^{2}}{(4\pi)^{2}} \frac{m_{t}^{2}}{M_{W}^{2}}(V_{\rm CKM})^{*}_{ts}(V_{\rm CKM})_{td}
\bigg [
\nonumber \\
& & 
\left \{
\frac{1}{4} \cdot  \frac{1}{D-4} \:
- \frac{3}{16}  + \frac{1}{8} {\rm log}\left ( \frac{m_{t}^{2}}{4\pi \mu ^{2}e^{-\gamma_{E}}} 
\right )
\right \} \gamma_{(\mu }(p+p^{\prime})_{\nu}L
\nonumber \\
& & 
+\left \{ - \frac{1}{2}\cdot  \frac{1}{D-4} 
+  \frac{1}{4} - \frac{1}{4} {\rm log}\left ( \frac{m_{t}^{2}}{4\pi \mu ^{2}e^{-\gamma_{E}}} 
\right ) 
+\frac{1}{4}
\right \} \eta_{\mu \nu} \left ( m_{s}L + m_{d}R \right )
\nonumber \\
& &
+{\cal O} \left (\frac{1}{m_{t}^{2}}  \right )
\bigg ]\:.
\label{eq:newgammamunucounterterm}
\end{eqnarray}
The fourth  term $`` \displaystyle{+\frac{1}{4}}" $ in the curly brackets 
in the third line of (\ref{eq:newgammamunucounterterm})   comes from the top quark contribution in 
the second term in (\ref{eq:deviation1}) and (\ref{eq:deviation2}).
It is  quite remarkable that there occurs a  cancellation among  the leading terms in 
(\ref{eq:largetopquarkmasslimitgammamunu}) and  those in  (\ref{eq:newgammamunucounterterm}) and  
the renormalized vertex  is not of the order of $m_{t}^{2}/M_{W}^{2}$ but of ${\cal O}(1)$, i.e., 
\begin{eqnarray}
& & \hskip-1cm
\Gamma_{\mu \nu}^{\rm ren} (p, p^{\prime})
=
\Gamma_{\mu \nu} (p, p^{\prime}) + \Gamma^{\rm c.t.}_{\mu \nu} (p, p^{\prime}) 
=
 {\cal O}\left ( 1 \right )\:.
\label{eq:renormalizedgammamunuorder1}
\end{eqnarray}
There is thus no enhancement by the factor $m_{t}^{2}/M_{W}^{2}$  
in the $d$-$s$-graviton vertex in the large top quark mass limit.

The cancellation between the leading terms in $\Gamma_{\mu \nu}$ and $\Gamma_{\mu \nu}^{\rm c.t.}$, 
however,  is not totally unexpected.  In fact we have seen in 
(\ref{eq:largetopquarkmasslimitgammamunu}) and  (\ref{eq:newgammamunucounterterm}) that 
the tensor-index-  and gamma-matrix-structures of 
$\Gamma_{\mu \nu}$ and $\Gamma_{\mu \nu}^{\rm c.t.}$ consist of two-types, i.e., 
$\gamma_{(\mu}(p+p^{\prime})_{\nu )}L$ 
and 
$\eta_{\mu \nu}(m_{s}L+m_{d}R)$.
Only with  these two types,   it is impossible  for $\Gamma_{\mu \nu}^{\rm ren}$
to satisfy the gravitational transverse condition 
(\ref{eq:transversalityequation}) on the mass-shell of external quarks. The sum  of the leading terms in 
$\Gamma_{\mu \nu}$ and $\Gamma_{\mu \nu}^{\rm c.t.}$ has necessarily to vanish. Note that 
in subleading orders, there appear several other types of tensor-index- and gamma-matrix-structures and 
the transverse condition would become non-trivial.

The absence of the ${\cal O}(m_{t}^{2}/M_{W}^{2})$ terms in $\Gamma_{\mu \nu}^{\rm ren}$ may  be 
seen  in terms of $\Sigma_{\rm ren}$ on the basis of the Ward-Takahashi identity.  
Let us now take the  large top quark mass limit in (\ref{eq:unrenormalisedsigma(p)}), i.e.,
\begin{eqnarray}
\Sigma (p)
& = &
\frac{g^{2}}{(4\pi)^{2}}(V_{\rm CKM})^{*}_{ts}(V_{\rm CKM})_{td} \frac{m_{t}^{2}}{M_{W}^{2}}
\bigg [
\nonumber \\
& &
\left \{
-\frac{1}{2}\cdot \frac{1}{D-4} - \frac{1}{4}{\rm log}\left (
\frac{m_{t}^{2}}{4\pi \mu^{2}e^{-\gamma_{E}}}\right ) + \frac{3}{8} \right \} \gamma \cdot p \: L
\nonumber \\
& & + \left \{
 \frac{1}{D-4} +\frac{1}{2}{\rm log}\left (
\frac{m_{t}^{2}}{4\pi \mu^{2}e^{-\gamma_{E}}}\right )  - \frac{1}{2}
\right \} (m_{s}L+m_{d}R) 
+ {\cal O} \left (  \frac{1}{m_{t}^{2}}   \right )\bigg ]\:.
\label{eq:approximatesigmap}
\end{eqnarray}
Then we combine  (\ref{eq:approximatesigmap}) with $\Sigma _{\rm c.t.}(p)$ in (\ref{eq:selfenergycounterterm}) 
with the four renormalization constants approximated by  
(\ref{eq:approximatezlXXX}), (\ref{eq:approximatezrXXX}), (\ref{eq:approximatezy1XXX}) and 
(\ref{eq:approximatezy2XXX}),   
\begin{eqnarray}
\hskip-0.5cm
\Sigma _{\rm c.t.} (p)
&=&
Z_{L} \gamma \cdot p \;L + Z_{R}\gamma \cdot p \:R + Z_{Y1}m_{s}L + Z_{Y2}m_{d}R\; 
\nonumber \\
&=&
\frac{g^{2}}{(4\pi)^{2}}(V_{\rm CKM})^{*}_{ts}(V_{\rm CKM})_{td} \frac{m_{t}^{2}}{M_{W}^{2}}
\bigg [
\nonumber \\
& &
\left \{
\frac{1}{2}\cdot \frac{1}{D-4} -\frac{3}{8} +  \frac{1}{4}{\rm log}\left (
\frac{m_{t}^{2}}{4\pi \mu^{2}e^{-\gamma_{E}}}\right ) 
 \right \} \gamma \cdot p \: L
\nonumber \\
& & + \left \{
 -  \frac{1}{D-4} + \frac{1}{2} - \frac{1}{2}{\rm log}\left (
\frac{m_{t}^{2}}{4\pi \mu^{2}e^{-\gamma_{E}}}\right )  
\right \} (m_{s}L+m_{d}R) 
+ {\cal O} \left (  \frac{1}{m_{t}^{2}}   \right )\bigg ]\:.
\label{eq:selfenergycountertermXXX}
\end{eqnarray}
Here we find the leading terms of ${\cal O}(m_{t}^{2}/M_{W}^{2})$ in  (\ref{eq:approximatesigmap}) 
and (\ref{eq:selfenergycountertermXXX}) cancelling each other,   and we 
 end  up with 
\begin{eqnarray}
\Sigma_{\rm ren} (p) &=&
\Sigma (p) + \Sigma _{\rm c.t.}(p) 
=
{\cal O}\left ( 1 \right )\:.
\label{eq:largetopmasssigmarenor}
\end{eqnarray}
The absence of  the ${\cal O}(m_{t}^{2}/M_{W}^{2})$ terms in 
$\Sigma_{\rm ren}(p)$
is consistent with the Ward-Takahashi identity (\ref{eq:wtidentityrenormalised}), 
whose left and right hand sides are both of ${\cal O}(1)$.


\section{The ${\cal O}(1)$ effective interactions }
 \label{sec:gravitationalpauliterm}

In the previous Section we   discussed seemingly most dominant 
terms behaving as  ${\cal O}(m_{t}^{2}/M_{W}^{2})$  
  when  the limit $m_{t} \to \infty $ is taken, and have shown that   these leading terms 
cancel among themselves.   Eqs. (\ref{eq:renormalizedgammamunuorder1}) and 
(\ref{eq:largetopmasssigmarenor}) were our net results in Section \ref{sec:effectivelagrangian}.
In the present Section we turn our attention to   
the  ${\cal O}(1)$ terms that are supposed to come next  
 in the said limit.
There are  a variety  of contributions to 
 this order  
and it is not  straightforward to 
classify all of them.  
For now   we simply highlight  a few   characteristic terms that are described  effectively by the operator 
 \begin{eqnarray}
& & \sqrt{-g}\: \overline \psi_{s} 
\left (m_{s}L + m_{d}R \right )\:  \psi_{d} \: {\cal R}
=
\sqrt{-e}\: \overline \Psi_{s} 
\left (m_{s}L + m_{d}R \right )\:  \Psi_{d} \: {\cal R}\:.
\label{eq:114}
\end{eqnarray}
Here the scalar curvature ${\cal R}$ should not be confused with the chiral projection $R$. 
The strange and down quark  fields  on the right hand side of 
(\ref{eq:114}) is given   weight $(-e)^{1/4}$ ($\Psi _{s}=(-e)^{1/4}\psi_{s}$, 
 $\Psi _{d}=(-e)^{1/4}\psi_{d}$) .

In the weak field approximation    as given in (\ref{eq:expansioninkappa})   we have 
 \begin{eqnarray}
 \sqrt{-g} \; {\cal R}
 &=&
 -2  \kappa  \left ( \partial^{\mu}\partial ^{\nu}-\eta^{\mu \nu} \partial ^{2} \right )
 \left (
 h_{\mu \nu} -\frac{1}{2}\eta_{\mu \nu} {h^{\lambda}}_{\lambda}
 \right ) + {\cal O}(\kappa ^{2})
 \nonumber \\
 &=&
-  2 \kappa  \left ( \partial ^{\mu } \partial ^{\nu} + \frac{1}{2}\partial ^{2} \eta^{\mu \nu} \right ) h_{\mu \nu}
 + {\cal O}(\kappa ^{2})
  \:,
  \label{eq:riemanntensorlinearterm}
 \end{eqnarray}
 and in the momentum space Eq. (\ref{eq:riemanntensorlinearterm}) becomes 
 \begin{eqnarray}
   2 \kappa  \left ( k^{\mu } k^{\nu} + \frac{1}{2}k^{2} \eta^{\mu \nu} \right ) h_{\mu \nu}
 + {\cal O}(\kappa ^{2})
  \: .
 \end{eqnarray}
 Here $k^{\mu}$ is the graviton momentum,  i.e.,
 $k^{\mu}=p^{\mu}-p^{\prime \: \mu}$. 
 Thus if we find  in $\Gamma_{\mu \nu} ( p, p^{\prime}) $   terms of the 
 following combination of tensor-index  and gamma-matrix structures
 \begin{eqnarray}
\left \{  \left (   p  -  p^{\prime} \right )_{\mu}   \left (   p  -  p^{\prime} \right )_{\nu}
 +\frac{1}{2} \left (   p  -  p^{\prime} \right )^{2}\eta_{\mu \nu}\right \}
 \left (  m_{s}L + m_{d}R   \right )\:, 
 \label{eq:structure}
 \end{eqnarray}
  then we are allowed to say that  these  
  terms are described effectively by the operator    (\ref{eq:114}).

Looking at the explicit results of   ${\cal G}_{\mu \nu}^{(X)}$ $ (X=a,b, \cdots , h)$ 
in Section \ref{sec:gravitationalflavorchangingvertex} closely,  we notice that only 
${\cal G}_{\mu \nu}^{(f)}$  and ${\cal G}_{\mu \nu}^{(h)}$ contain terms that could possibly 
be given  the structure of  (\ref{eq:structure}):
\begin{eqnarray}
{\cal G}_{\mu \nu}^{(f)}
&=&  
 \frac{\kappa g^{2}}{(4\pi )^{2}} \frac{1}{M_{W}^{2}}\bigg [
G_{7}(p, p^{\prime})\eta_{\mu \nu}
\nonumber \\
& & 
+\left \{
f_{13}(p, p^{\prime})p_{\mu}^{\prime}p_{\nu}^{\prime}+ f_{13}(p^{\prime}, p)p^{}_{\mu}p^{}_{\nu}
- f_{14}(p, p^{\prime}) p_{(\mu}p_{\nu )}^{\prime}
\right \} \bigg ] m_{j}^{2}(m_{s}L+m_{d}R)
\nonumber \\
& & + \cdots \cdots \:, 
\label{eq:possiblef}
\end{eqnarray}
\begin{eqnarray}
{\cal G}_{\mu \nu}^{(h)}
&=&
\frac{\kappa g^{2}}{(4\pi )^{2}} \frac{1}{M_{W}^{2}}\bigg [
G_{12}(p, p^{\prime})\eta_{\mu \nu}
\nonumber \\
& & 
+\left \{
\frac{1}{2}f_{24}(p, p^{\prime})p_{\mu}p_{\nu}+
 \frac{1}{2}f_{24}(p^{\prime}, p)p^{\prime}_{\mu}p^{\prime}_{\nu}
+ \frac{1}{2} f_{23}(p, p^{\prime}) p_{(\mu}p_{\nu )}^{\prime}
\right \} 
\bigg ] m_{j}^{2}(m_{s}L+m_{d}R)
\nonumber \\
& & + \cdots \cdots \:.
\label{eq:possibleh}
\end{eqnarray}
In order to confirm that terms in (\ref{eq:possiblef}) and (\ref{eq:possibleh}) are actually combined 
together to be given  the structure of   (\ref{eq:structure}),  we restrict our analyses to the 
following low energy case, 
\begin{eqnarray}
p^{2}, \: p^{\prime 2}, \: (p-p^{\prime})^{2} \ll M_{W}^{2}, \: m_{j}^{2}\:.
\label{eq:lowenergyapproximation}
\end{eqnarray}
Note that we do not assume any particular relation between $M_{W}$ and $m_{j} \: (j=t,c,u)$.

Applying the 
approximation (\ref{eq:lowenergyapproximation}) to the quantities  
$f_{13}(p, p^{\prime})$ and $f_{14}(p, p^{\prime})$ in (\ref{eq:possiblef}), 
and to 
$f_{23}(p, p^{\prime})$ and  $f_{24}(p, p^{\prime})$ in 
 (\ref{eq:possibleh}), we 
  just  set $p^{2}=p^{\prime \:2}=(p-p^{\prime})^{2}=0$ 
 in the integral representations (\ref{eq:f13}), (\ref{eq:f14}), (\ref{eq:f23}) and (\ref{eq:f24}).
 After performing double integration we get the following  formulae for the two combinations 
 of these functions
\begin{eqnarray}
f_{13}(0, 0) + \frac{1}{2} f_{24}(0, 0) &=& \frac{1}{m_{j}^{2}} F_{1} \left ( \frac{m_{j}^{2}}{M_{W}^{2}} 
\right )\:,
\label{eq:keisuuofpmupnu}
\\
f_{14}(0, 0) - \frac{1}{2} f_{23}(0, 0) &=& \frac{2}{m_{j}^{2}} F_{1} \left ( \frac{m_{j}^{2}}{M_{W}^{2}} 
\right )\:,
\label{eq:keisuuofpmupnuprime}
\end{eqnarray}
where we have introduced a  function
\begin{eqnarray}
F_{1} (x)=\frac{x(3-x)}{8(1-x)^{2}} - \frac{x(2 x^{2}-4x-1)}{12(1-x)^{3}} \: {\rm log} \: x \:\:.
\end{eqnarray}
Note that the function $F_{1}(x)$ is finite at $x=1$, i.e., 
$\lim _{x \to 1} F_{1}(x)=1/12$. 
Also note the asymptotic behavior, 
$\displaystyle{F_{1}(x) \sim - \frac{1}{8} + \frac{1}{6}\:{\rm log}\:x}$
for large $x$.
It is remarkable that 
a common quantity $F_{1}(m_{j}^{2}/M_{W}^{2})$ has appeared on the right hand side of 
(\ref{eq:keisuuofpmupnu}) and (\ref{eq:keisuuofpmupnuprime}). 
Thanks to this common quantity, the sum of all the terms with 
$p_{\mu}p_{\nu}$, $p^{\prime}_{\mu}p^{\prime}_{\nu}$ and $p_{(\mu}p^{\prime}_{\nu )}$ 
  in (\ref{eq:possiblef}) and (\ref{eq:possibleh})  turns out to be a very concise one, i.e.,
\begin{eqnarray}
& & \hskip-0.5cm 
\left \{ f_{13}(0, 0) + \frac{1}{2} f_{24}(0, 0)  \right \} (p_{\mu }p_{\nu} + p^{\prime}_{\mu}p^{\prime}_{\nu})
-
\left \{ f_{14}(0, 0) - \frac{1}{2} f_{23}(0, 0) \right \} p_{(\mu}p^{\prime}_{\nu )}
\nonumber \\ 
& & \hskip1cm =
 \frac{1}{m_{j}^{2}} F_{1} \left ( \frac{m_{j}^{2}}{M_{W}^{2}} \right )
  (p-p^{\prime})_{\mu} (p-p^{\prime})_{\nu}\:.
  \label{eq:conciseone}
\end{eqnarray}

Let us now move to the remaining terms, $G_{7}(p, p^{\prime})\eta_{\mu \nu}$ in 
(\ref{eq:possiblef}) and $G_{12}(p, p^{\prime})\eta_{\mu \nu}$ in  (\ref{eq:possibleh}).
Recall that $G_{7}(p, p^{\prime})$ contains $f_{4}(p, p^{\prime})$, $f_{6}(p, p^{\prime})$ and 
$f_{15}(p, p^{\prime})$ as defined in (\ref{eq:G7ppprime}) and that 
$G_{12}(p, p^{\prime})$ contains $f_{20}(p, p^{\prime})$, $f_{24}(p, p^{\prime})$, 
$f_{26}(p, p^{\prime})$ and 
$f_{27}(p, p^{\prime})$ as defined in (\ref{eq:G12ppprime}).
We expand these functions in Taylor series with respect to $p^{2}$, $p^{\prime 2}$ and 
$(p - p^{\prime})^{2}$ through the second order to meet with (\ref{eq:conciseone}). 
After straightforward calculations we have found a formula
\begin{eqnarray}
G_{7}(p, p^{\prime})+G_{12}(p, p^{\prime})
&=&
G_{7}(0, 0)+G_{12}(0, 0) + \frac{(p^{2}+p^{\prime \: 2})}{m_{j}^{2}} F_{2} \left ( \frac{m_{j}^{2}}{M_{W}^{2}} \right )
\nonumber \\
& & + \frac{(p-p^{\prime})^{2}}{m_{j}^{2}}\cdot \frac{1}{2} F_{1}\left ( \frac{m_{j}^{2}}{M_{W}^{2} } \right )
+ \cdots \cdots \:,
\label{eq:g7plusg12}
\end{eqnarray}
where the ellipses denote higher order terms in the Taylor expansion and are neglected. 
Here we have defined another  function 
\begin{eqnarray}
F_{2}(x)=\frac{x + x^{2}}{8(1-x)^{2}}+\frac{x^{2}}{4(1-x)^{3}}\:{\rm log}\:x\:.
\end{eqnarray}
This function is also free from singularity at $x=1$, i.e., 
 $\lim _{x \to 1} F_{2}(x)=1/24$. 
The third term in (\ref{eq:g7plusg12}) that contains this function $F_{2}(m_{j}^{2}/ M_{W}^{2})$ would be 
described by an operator of a different type from (\ref{eq:114}), and we will not delve into it hereafter.
It is noteworthy that the quantity $F_{1}(m_{j}^{2}/M_{W}^{2})$ has again appeared as the coefficient of $(p-p^{\prime})^{2}$ in (\ref{eq:g7plusg12}).

Those related to the graviton momentum 
$(p-p^{\prime})^{\mu}$ are thus 
summed up with the common coefficient $F_{1}(m_{J}^{2}/M_{W}^{2})$ as
\begin{eqnarray}
{\cal G}_{\mu \nu}^{(f)}+ {\cal G}_{\mu \nu}^{(h)}
&=&
\frac{\kappa g^{2}}{(4 \pi)^{2}}
 \frac{1}{M_{W}^{2}} F_{1} \left ( \frac{m_{j}^{2}}{M_{W}^{2}} \right )
 \left \{
  (p-p^{\prime})_{\mu} (p-p^{\prime})_{\nu} + \frac{1}{2} (p-p^{\prime})^{2} \eta_{\mu \nu}
  \right \}
  \nonumber \\
  & & \hskip5cm    \times (m_{s}L + m_{d}R)
  \nonumber \\
  & & + \cdots \cdots \: .
  \end{eqnarray}
In terms of $\Gamma_{\mu \nu}(p, p^{\prime})$, we have 
\begin{eqnarray}
\Gamma_{\mu \nu}( p, p^{\prime})
&=&
\frac{\kappa g^{2}}{(4 \pi)^{2}}
 \frac{{\cal F}_{1}}{M_{W}^{2}}
\left \{
  (p-p^{\prime})_{\mu} (p-p^{\prime})_{\nu} + \frac{1}{2} (p-p^{\prime})^{2} \eta_{\mu \nu}
  \right \}
   (m_{s}L + m_{d}R)
  \nonumber \\
  & & + \cdots \cdots \;,
  \label{eq:1018}
  \end{eqnarray}
  where the coefficient in front of the brackets 
  \begin{eqnarray}
  {\cal F}_{1}&=& \sum_{j=t,c,u}
 (V_{\rm CKM})_{js}^{*}(V_{\rm CKM})_{jd}
  F_{1} \left ( \frac{m_{j}^{2}}{M_{W}^{2}} \right )
  \label{eq:thecoefficientsinfrontofthebrackets}
  \end{eqnarray}
  depends on the top, charm and up quark masses  
  as well as the CKM matrix elements.   Eq. (\ref{eq:1018}) is given  the same 
  tensor-index and gamma-matrix structure as  (\ref{eq:structure}).   This  
is exactly what we expect to arise from the operator (\ref{eq:114}), and 
the effective Lagrangian  becomes  
  \begin{eqnarray}
{\cal L}_{\rm eff} ^{{\cal R}}
&=&
\frac{g^{2}}{(4\pi)^{2}}\frac{{\cal F}_{1}}{2 M_{W}^{2}}
\sqrt{-e}\: \overline \Psi_{s} 
\left (m_{s}L + m_{d}R \right )\:  \Psi_{d} \: {\cal R}.
\label{eq:importantformula}
\end{eqnarray}
As we remarked before, the function $F_{1}$ has  the asymptotic behavior 
\begin{eqnarray}
F_{1} \left ( \frac{m_{j}^{2}}{M_{W}^{2}} \right ) \sim -\frac{1}{8} +
\frac{1}{6} {\rm log}\left ( \frac{m_{j}^{2}}{M_{W}^{2}}  \right ) \:\:\:\: {\rm as } \:\:\:\: 
\frac{m_{j}^{2}}{M_{W}^{2}} \to \infty \:, 
\end{eqnarray}
and this formula shows clearly the ${\cal O}(1)$   
non-decoupling effects  of the heavy quark.
Numerically,  the top quark contribution  to the coefficient ${\cal F}_{1}$ 
is the  most dominant over the other two, as we find  
\begin{eqnarray}
& &  F_{1}\left ( \frac{m_{t}^{2}}{M_{W}^{2}} \right )=0.21686\; ,  
\\
& & F_{1} \left ( \frac{m_{c}^{2}}{M_{W}^{2}} \right ) =-7.92 \times 10^{-5} \:, 
 \\
& & F_{1}\left ( \frac{m_{u}^{2}}{M_{W}^{2}} \right )=
-9.96 \times 10^{-10} \:,
\label{eq:f1equal021686}
\end{eqnarray}
for $M_{W}=80.379$ GeV,  $m_{t}=172.76\:\:{\rm GeV}$, $m_{c}=1.27\:\:\:{\rm GeV}$
and $m_{u}=2.16 \:\:\:{\rm MeV}$    \cite{pdg}. 
This non-negligible effect of the heavy top quark is  a manifestation  of the 
${\cal O}(1)$    non-decoupling effects.

Although our effective Lagrangian (\ref{eq:importantformula}) is  one of the most important  results 
of the present paper, we do not attempt here to apply  (\ref{eq:importantformula}) 
  to actual physical problems. Let us, however, bear in our mind thtat
 (\ref{eq:importantformula})  could be  relevant to flavor-changing and  CP-violating gravitational phenomena. In fact, the most dominant top quark contribution 
in (\ref{eq:thecoefficientsinfrontofthebrackets})  is accompanied by 
$(V_{\rm CKM})_{ts}^{*}(V_{\rm CKM})_{td}$ which is  given by 
\begin{eqnarray}
(V_{\rm CKM})_{ts}^{*}(V_{\rm CKM})_{td}=
\left (
-c_{12}s_{23}-s_{12}c_{23}s_{13}e^{i\delta}
\right )^{*}
\left (
s_{12}s_{23}-c_{12}c_{23}s_{13}e^{i\delta}
\right )\:,
\end{eqnarray}
according to the standard parametrization  \cite{pdg},  and 
 contains the CP-violating phase $\delta$. 

Finally we would like to add a comment on the comparison with the loop-induced 
$d \to  s + \gamma $ transition, on  which it has been 
 pointed out  \cite{inamilim, deshpande1} 
 that the transition amplitude contains  a term 
described effectively by the operator  
\begin{eqnarray}
\overline \psi_{s}\sigma^{\mu \nu}\left (m_{s}L + m_{d}R \right )\psi_{d} \: F_{\mu \nu}\:.
 \label{eq:abelianpaulioperator}
\end{eqnarray}
Here $F_{\mu \nu}$ is the 
electromagnetic field strength  and $\sigma ^{\mu \nu}$ is defined in (\ref{eq:sigmaab}). 
This operator reminds us of  the Pauli term in quantum electrodynamics. 
It has also been known   \cite{eilam, hou, deshpande3}
 that in the loop-induced  $d \to s + {\rm gluon}$ transition,   there also exist 
contributions described by the similar operator 
\begin{eqnarray}
 \overline \psi_{s} \: T^{a}\: \sigma^{\mu \nu}\left (m_{s}L + m_{d}R \right )\:  \psi_{d} \: F^{a}_{\mu \nu}\:,
 \label{eq:nonabelianpaulioperator}
\end{eqnarray}
where $F^{a}_{\mu \nu}$ is the field strength of the gluon field and $T^{a}$ is the generator of the 
 color gauge group. 

A question naturally arises here: one may ask whether there exists  a similar sort of contribution 
 in the gravitational 
 process (\ref{eq:dsgraviton}).   It is very tempting to postulate  that the operator 
 analogous to (\ref{eq:nonabelianpaulioperator}) would be 
\begin{eqnarray}
& & \sqrt{-g}\: \overline \psi_{s} \left \{ \sigma ^{ab}, \: \sigma^{\mu \nu} \right \} 
\left (m_{s}L + m_{d}R \right )\:  \psi_{d} \: R_{\mu \nu a b }\:.
\label{eq:gravitationalpauliterm}
\end{eqnarray}
Here the non-abelian field strength $F^{a}_{\mu \nu} $ in  (\ref{eq:nonabelianpaulioperator})  is 
replaced by the Riemann tensor defined   in terms of the spin connection as 
\begin{eqnarray}
R_{\mu \nu ab}
&=&
\partial _{\mu}  \omega_{\nu ab}
- \partial _{\nu} \omega_{\mu ab}
+
  {\omega_{\mu a}}^{c}  \omega_{\nu c b}
-
{\omega_{\nu a}}^{c} \omega_{\mu c b}
\:\:\: \left ( ={e^{\lambda}}_{a} {e^{\rho}}_{b} R_{\mu \nu \lambda \rho}\right )
\:.
\end{eqnarray}
The gauge group generator $T^{a}$ in  (\ref{eq:nonabelianpaulioperator}) has been 
 replaced by the local Lorentz group generator $\sigma^{ab}$.

 Now it is known that the Riemann tensor may be decomposed into three parts
 \begin{eqnarray}
R_{\lambda \mu \nu \rho}
&=&
C_{\lambda \mu \nu \rho} - \frac{1}{2}\left (
R_{\lambda \rho}g_{\mu \nu}
-
R_{\lambda \nu}g_{\mu \rho}
+
R_{\mu \nu}g_{\lambda \rho}
-
R_{\mu \rho}g_{\lambda \nu}
\right )
 - \frac{1}{6}
{\cal R}\: \left (
g_{\lambda \nu} g_{\mu \rho } - g_{\lambda \rho } g_{\mu \nu}
\right )\:, 
\nonumber \\
\label{eq:decomposition}
\end{eqnarray}
where the first term $C_{\lambda \mu \nu\rho} $  is the Weyl tensor and is traceless 
\begin{eqnarray}
g^{\lambda \nu}C_{\lambda \mu \nu \rho}=0, \:\:\:\:
g^{\mu \rho}C_{\lambda \mu \nu \rho}=0\:.
\end{eqnarray}
Once we take the product of (\ref{eq:decomposition}) with 
$\{ \sigma^{\lambda \mu},  \sigma ^{\nu \rho} \}$, 
we immediately find a relation 
 \begin{eqnarray}
\left \{   \sigma ^{\lambda \mu} , \sigma ^{ \nu \rho} \right \}\:R_{\lambda \mu \nu \rho}
=
4{\cal R}  + \left \{   \sigma ^{\lambda \mu} , \sigma ^{ \nu \rho} \right \}\:C_{\lambda \mu \nu \rho}\:.
\label{eq:pauli222}
\end{eqnarray}
The scalar curvature term $4{\cal R}$ on the right hand side of (\ref{eq:pauli222}), 
when plugged into (\ref{eq:gravitationalpauliterm}), gives us the same operator as 
in (\ref{eq:114}), which has already been studied above.
It is therefore crucial whether the contribution due to 
$\{ \sigma^{\lambda \mu}, \sigma^{\nu \rho} \} C_{\lambda \mu \nu \rho}$ 
exists or not in the amplitudes in order for the operator 
(\ref{eq:gravitationalpauliterm})  to be an effective one. 
Unfortunately in the weak field expansion (\ref{eq:expansioninkappa}),  a straightforward 
calculation shows 
 \begin{eqnarray}
\left \{   \sigma ^{\lambda \mu} , \sigma ^{ \nu \rho} \right \}\:R_{\lambda \mu \nu \rho}
-
4{\cal R} 
=
 {\cal O}(\kappa^{2})  \:.
\end{eqnarray}
This means that  the Weyl tensor contribution 
$\{ \sigma^{\lambda \mu}, \sigma^{\nu \rho} \} C_{\lambda \mu \nu \rho}$ 
 is  of   the order of $\kappa^{2}$  
 and cannot be seen in our ${\cal O}(\kappa )$ calculation.  To seek for a gravitational analogue of 
 (\ref{eq:abelianpaulioperator}) and (\ref{eq:nonabelianpaulioperator}), we have to examine 
 two graviton emission processes.


\section{Summary}
\label{sec:summary}

In the present paper we have investigated the loop-induced flavor-changing 
gravitational process (\ref{eq:dsgraviton}) in the standard electroweak theory
in order to see the non-decoupling effects of the heavy top quark running along an internal line.
We have confirmed explicitly that the renormalization constants $Z_{L}$, $Z_{R}$, $Z_{Y1}$ and 
$Z_{Y2}$ determined  for the self-energy type $d \to s$ diagrams (Figure \ref{fig:exercise2}) in flat space serve adequately to eliminate ultraviolet divergences in the one-graviton vertex diagrams 
(Figure \ref{fig:attachedtovertex} and Figure \ref{fig:attachedtopropagators}).  It is pointed out 
that the unrenormalized and renormalized two- and three-point functions satisfy the same form 
of Ward-Takahashi identities, 
 (\ref{eq:wtidentitybare}) and (\ref{eq:wtidentityrenormalised}), 
 similarly to  quantum electrodynamics. 
We collected and  examined the leading terms in the $m_{t} \to \infty $ limit 
in the renormalized transition amplitude that are proportional to $m_{t}^{2}/M_{W}^{2}$. 
We have found that these ${\cal O}(m_{t}^{2}/M_{W}^{2})$ terms disappear by cancellation. 
The non-decoupling effects of the internal  top quark  thus take place at the ${\cal O}(1)$ 
level.
Among the 
${\cal O}(1)$
   terms, we have noticed  the contributions which are supposed to have come 
from the effective Lagrangian   (\ref{eq:importantformula}) that consists of 
quark bilinear form coupled to the space-time scalar curvature $\cal R$. 
The top quark effect is sizable in 
(\ref{eq:importantformula}) and this is 
one of manifiest    forms
 of non-decoupling effects.

While   the effective Lagrangian   (\ref{eq:importantformula})  looks concise, 
we did not find  the Ricchi tensor  or the Weyl tensor counterpart within the  present standard model  calculation of  (\ref{eq:dsgraviton}) at the one-loop and one-graviton emission level. 
Perhaps in more sophisticated models such as supersymmetric gauge theories or 
 grand unification models,
 in which several very heavy particles are supposed to exist,  
we could encounter various types of effective interactions  
as explored extensively  by Ruhdorfer et al \cite{ruhdorfer}.
Or such various interaction terms would arise in two-loop or higher level of calculations. 
Those non-trivial effective interactions with spacetime could cause  intriguing effects 
if applied to the early universe. When the universe was expanding, the Riemann tensor, Ricchi tensor 
and scalar curvature in the ensuing effective Lagrangian   have to be those of 
the Friedmann-Lema{\^ i}tre-Robertson-Walker metric,  and the effective interactions among  quarks 
would not respect the time-reversal invariance. Implications of such effective interactions would be 
extremely interesting and deserve further pursuit, 
but for now we have to leave these investigations for our future work.


\section*{Acknowledgment}
The authors wish to thank Professor T.  Kugo,  Professor C. S. Lim 
and Professor K. Izumi for stimulating  discussions. 
They are also grateful to Dr. Shu-Yu Ho for calling their attention to  early references.
Last but not least the authors would like to express their sincere thanks to the anonymous referee 
for her/his  penetrating comments that were very useful to improve the present paper.

\appendix
\section{The  Feynman parameters' integrations}
\label{sec:integralrepresentation1}

The following parameter integrations appear in evaluating the self-energy type $d \to s $ transition amplitudes
\begin{eqnarray}
f_{1}( p^{2})
&=&\int _{0}^{1} dx \: 
(1-x) {\rm log} \left \{
\frac{-x(1-x)p^{2}+xm_{j}^{2}+(1-x)M_{W}^{2}}{4\pi \mu^{2}e^{-\gamma_{E}}}
\right \}\;, 
\label{eq:f1(pp)}
\\
f_{2}( p^{2})
&=&\int _{0}^{1} dx \: 
 {\rm log} \left \{
\frac{-x(1-x)p^{2}+xm_{j}^{2}+(1-x)M_{W}^{2}}{4\pi \mu^{2}e^{-\gamma_{E}}}
\right \}\;, 
\label{eq:f2(pp)}
\end{eqnarray}
where $\gamma_{E}$ is the Euler number.  These functions appear also in the 
calculation of Figure \ref{fig:attachedtovertex}.  Note that both (\ref{eq:f1(pp)}) and (\ref{eq:f2(pp)}) 
are $m_{j}$-dependent, although the dependence is not made explicit on the left hand side of 
(\ref{eq:f1(pp)}) or (\ref{eq:f2(pp)}). 
The same comment applies to all the functions to be introduced hereafter  in this Appendix. 

Combining propagators in Figure \ref{fig:attachedtopropagators}  (e) and   
Figure \ref{fig:attachedtopropagators}    (f) by using Feynman's parameters, 
the following combination commonly appears in the denominator:
\begin{eqnarray}
\Delta_{1}
&\equiv &
-y(1-x-y)p^{2} -x(1-x-y)p^{\prime\:2} -xy (p-p^{\prime})^{2}
\nonumber \\
& & +(x+y)M_{W}^{2}+(1-x-y)m_{j}^{2}\:.
\label{eq:deltaone}
\end{eqnarray}
The parameter integrations involving (\ref{eq:deltaone}) 
that we used in Section  \ref{sec:gravitationalflavorchangingvertex}  are as follows:

\begin{eqnarray}
f_{3}(p, p^{\prime})&=&M_{W}^{2}\:\int _{0}^{1}dx \int_{0}^{1-x}dy\: \frac{y}{\Delta_{1}}\:,
\\
f_{4}(p, p^{\prime})&=&\int _{0}^{1}dx \int_{0}^{1-x}dy\: y\: {\rm log}\left \{
\frac{\Delta_{1}}{4\pi \mu^{2}e^{-\gamma_{E}}}
\right \}\:,
\\
f_{5}(p, p^{\prime})&=&
M_{W}^{2}\int _{0}^{1}dx \int _{0}^{1-x}dy\: \frac{y(x+y)}{\Delta_{1}}\:,
\\
f_{6}(p, p^{\prime})&=&\int _{0}^{1}dx \int_{0}^{1-x}dy\: (1-4y) \: {\rm log}\left \{
\frac{\Delta_{1}}{4\pi \mu^{2}e^{-\gamma_{E}}}
\right \}\:,
\\
f_{7}(p, p^{\prime})&=&\int _{0}^{1} dx \int _{0}^{1-x} dy \: \frac{y^{2}(1-y)}{\Delta_{1}}\:,
\\
f_{8}(p, p^{\prime})&=&\int _{0}^{1} dx \int _{0}^{1-x} dy \: \frac{x^{2}(1+y)}{\Delta_{1}}\:,
\\
f_{9}(p, p^{\prime})&=&\int _{0}^{1} dx \int _{0}^{1-x} dy \: \frac{x(2y^{2} -1)}{\Delta_{1}}\:,
\\
f_{10}(p, p^{\prime})&=&\int _{0}^{1}dx \int _{0}^{1-x}dy \: \frac{(x+y)(1-2y)}{\Delta_{1}}\:,
\\
f_{11}(p, p^{\prime})&=&\int_{0}^{1}dx \int_{0}^{1-x}dy\: \frac{xy(1-y)}{\Delta_{1}}\:,
\\
f_{12}(p, p^{\prime})&=&\int_{0}^{1}dx \int_{0}^{1-x}dy\: \frac{x(1-x-y+2xy)}{\Delta_{1}}\:,
\\
f_{13}(p, p^{\prime})&=&\int_{0}^{1}dx \int_{0}^{1-x}dy\;\frac{x(1-x)}{\Delta_{1}}\:,
\label{eq:f13}
\end{eqnarray}
\begin{eqnarray}
f_{14}(p, p^{\prime})&=&\int_{0}^{1}dx \int_{0}^{1-x} dy\: \frac{(1-x-y+2xy)}{\Delta_{1}}\:,
\label{eq:f14}
\\
f_{15}(p, p^{\prime})&=& M_{W}^{2} \int_{0}^{1}dx \int_{0}^{1-x}dy\: \frac{1}{\Delta_{1}}\:.
\end{eqnarray}

Similarly when we combine  propagators in Figure \ref{fig:attachedtopropagators}  (g) and   
Figure \ref{fig:attachedtopropagators}    (h) by using Feynman's parameters,  
the denominator turns out to be 
\begin{eqnarray}
\Delta_{2}&\equiv&-y(1-x-y)p^{2}-x(1-x-y)p^{\prime \: 2} - xy (p-p^{\prime})^{2}
\nonumber \\
& & +(x+y)m_{j}^{2}+(1-x-y)M_{W}^{2}\:.
\end{eqnarray}
The parameter integrations containing $\Delta_{2}$ are listed below:
\begin{eqnarray}
f_{16}(p, p^{\prime})
&=&\int _{0}^{1}dx \int _{0}^{1-x}dy\:(1-2y)\:{\rm log}\left \{
\frac{\Delta_{2}}{4\pi \mu^{2} e^{-\gamma_{E}}}
\right \}\:,
\\
f_{17}(p, p^{\prime})
&=&\int _{0}^{1}dx \int _{0}^{1_x}dy\:\frac{y(1-y)(1-2y)}{\Delta_{2}} \:,
\\
f_{18}(p, p^{\prime})
&=&\int _{0}^{1}dx \int _{0}^{1-x}dy\: \frac{x(1-x)(1-2y)}{\Delta_{2}}\:,
\\
f_{19}(p, p^{\prime})
&=&\int _{0}^{1}dx \int _{0}^{1-x}dy\: \frac{(1-x)(1-y)(1-2y)}{\Delta_{2}}\:,
\\
f_{20}(p, p^{\prime})
&=&\int _{0}^{1}dx \int _{0}^{1-x}dy\: \frac{1-2y}{\Delta_{2}}\:,
\\
f_{21}(p, p^{\prime})
&=&\int _{0}^{1}dx \int _{0}^{1-x}dy\: \frac{(1-y)(1-x-3y + 4xy)}{\Delta_{2}}\:,
\\
f_{22}(p, p^{\prime})
&=&\int _{0}^{1}dx \int _{0}^{1-x}dy\: \frac{(1-2y)(1-x-y)}{\Delta_{2}}\:,
\\
f_{23}(p, p^{\prime})&=&\int _{0}^{1}dx \int _{0}^{1-x}dy\: \frac{(x+y-4xy) }{\Delta_{2}}\:,
\label{eq:f23}
\\
f_{24}(p, p^{\prime})&=&\int _{0}^{1}dx \int _{0}^{1-x}dy\: \frac{ y(1-2y) }{\Delta_{2}}\:,
\label{eq:f24}
\\
f_{25}(p, p^{\prime})&=&\int _{0}^{1}dx \int _{0}^{1-x}dy\: \frac{( 1 - x - y) }{\Delta_{2}}\:,
\\
f_{26}(p, p^{\prime})&=&\int _{0}^{1}dx \int _{0}^{1-x}dy\: \frac{ y( 1 - y) }{\Delta_{2}}\:,
\\
f_{27}(p, p^{\prime})&=&\int _{0}^{1}dx \int _{0}^{1-x}dy\: \frac{ xy }{\Delta_{2}}\:,
\\
f_{28}(p, p^{\prime})&=&\int _{0}^{1}dx \int _{0}^{1-x}dy\: \frac{ xy(1-2y) }{\Delta_{2}}\:,
\\
f_{29}(p, p^{\prime})&=&\int _{0}^{1}dx \int _{0}^{1-x}dy\: \frac{  y(1-3x-y+4xy) }{\Delta_{2}}\:.
\end{eqnarray}

\vfill\eject
\section{Functions $G_{i}$ ($i=1, \cdots , 12$) }
\label{appendixb}
Some combinations of Feynman parameters' integrations are defined below : 
\begin{eqnarray}
G_{1}(p^{2})
&=&
\frac{1}{D-4} +f_{1}(p^{2})\:,
\label{eq:g1pp}
\\
G_{2}(p^{2})
&=&
\frac{2}{D-4} +f_{2}(p^{2})\:,
\label{eq:g2pp}
\\
G_{3}(p, p^{\prime})
&=&\frac{1}{3}\cdot \frac{1}{D-4} + \frac{1}{6} -f_{3}(p, p^{\prime}) +f_{4}(p, p^{\prime})
\:,
\\
G_{4}(p, p^{\prime})
&=&
-\frac{4}{3}\cdot \frac{1}{D-4} - \frac{5}{6} +\left (
2-\frac{p^{2}}{M_{W}^{2}} +\frac{2m_{j}^{2}}{M_{W}^{2}}
\right ) f_{3}(p, p^{\prime}) +\frac{p^{\prime \: 2}}{M_{W}^{2}} f_{3}(p^{\prime} , p )
\nonumber \\
& & -4\:f_{4}(p, p^{\prime}) + 2\left (
1-\frac{m_{j}^{2}}{M_{W}^{2}}
\right ) f_{5}(p, p^{\prime}) \;,
\\
G_{5}(p, p^{\prime})
&=&
\frac{1}{6}\cdot \frac{1}{D-4} -\frac{1}{2} f_{3}(p, p^{\prime}) + \frac{1}{2} f_{4}(p, p^{\prime})\:,
\\
G_{6}(p, p^{\prime})
&=&
- \frac{1}{6}\cdot \frac{1}{D-4} -\frac{1}{2} f_{6}(p, p^{\prime}) -  f_{4}(p, p^{\prime})\:,
\\ 
G_{7}(p, p^{\prime})
&=&
- \frac{1}{2}\cdot \frac{1}{D-4} -\frac{1}{2} f_{6}(p, p^{\prime}) -  2f_{4}(p, p^{\prime})
+\frac{1}{2}f_{15}(p, p^{\prime}) \:,
\label{eq:G7ppprime}
\\
G_{8}(p, p^{\prime})
&=&
\frac{1}{6} \cdot \frac{1}{D-4} + \frac{1}{12} + \frac{1}{2}f_{16}(p, p^{\prime}) 
+ m_{j}^{2}f_{20}(p, p^{\prime})\:,
\\
G_{9}(p, p^{\prime})
&=&
-\frac{1}{6} \cdot \frac{1}{D-4} - \frac{1}{6} - \frac{1}{2}f_{16}(p, p^{\prime}) 
- p^{2}f_{17}(p, p^{\prime})  -  p^{\prime\: 2} f_{18}(p, p^{\prime})
\nonumber \\
& & + 2 \: p\cdot p^{\prime}\: f_{19}(p, p^{\prime})  - m_{j}^{2}f_{20}(p, p^{\prime})\:,
\\
G_{10}(p, p^{\prime})
&=&
\frac{1}{12} \cdot \frac{1}{D-4}  + \frac{1}{4}f_{16}(p, p^{\prime}) 
- \frac{1}{4}m_{j}^{2}f_{20}(p, p^{\prime})\:,
\\
G_{11}(p, p^{\prime})
&=&
-\frac{1}{12}\cdot \frac{1}{D-4} - \frac{1}{24} - \frac{1}{4}f_{16}(p, p^{\prime}) 
+\frac{1}{4}p^{2}f_{17}(p, p^{\prime}) + \frac{1}{4}p^{\prime\:2}f_{18}(p, p^{\prime}) 
\nonumber 
\\
& & 
-\frac{1}{2}p\cdot p^{\prime} f_{28}(p, p^{\prime}) +\frac{1}{4}m_{j}^{2}f_{20}(p, p^{\prime})\:,
\\
G_{12}(p, p^{\prime})
&=&
- \frac{1}{8} - \frac{1}{4}p^{2} f_{26}(p, p^{\prime}) -\frac{1}{4}p^{\prime \:2} f_{26}(p^{\prime}, p)
+\frac{1}{2}p\cdot p^{\prime} f_{27}(p, p^{\prime})
\nonumber \\
& & +m_{j}^{2}\left \{
\frac{1}{4}f_{20}(p, p^{\prime}) -\frac{1}{2}f_{24}(p, p^{\prime}) + f_{26}(p, p^{\prime})
\right \}\:.
\label{eq:G12ppprime}
\end{eqnarray}


%

\vspace{0.2cm}
\noindent


\let\doi\relax


\begin{thebibliography}{9}

\bibitem{abbott0}
B.P. Abbott et al.,
\PRL{116,061102,2016}

\doi{https://doi.org/10.1103/PhysRevLett.116.061102}
%
%
\bibitem{abbott1}
B.P. Abbott et al.,
\PRL{116,241103,2016}

\doi{https://doi.org/10.1103/PhysRevLett.116.241103}
%
%
\bibitem{abbott2}
B.P. Abbott et al.,  
\PRL{118,0221101,2017}

\doi{https://doi.org/10.1103/PhysRevLett.118.221101}
%
\bibitem{abbott3}
B.P. Abbott et al.,  
\PRL{119,161101,2017}

\doi{https://doi.org/10.1103/PhysRevLett.119.161101}
%

%
%
\bibitem{nanograv}
Z. Arzoumanian et al., 
Astrophys. J. Letters {\bf 905}, L34   (2020)    [arXiv: 2009.04496 [astro-ph.HE]].
 
\doi{https://doi.org/10..3847/2041-8213/abd401}
%
\bibitem{lommen}
A.N. Lommen, 
Rept. Prog. Phys. {\bf 78}, 124901 
(2015).

\doi{https://doi.org/10.1088/0034-4885/78/12/124901}
%
\bibitem{tiburzi}
C. Tiburzi, 
Publ. Astron. Soc. Austral. {\bf 35}, e013  (2018) [arXiv: 1802.05076 [astro-ph.IM]]. 

\doi{https://doi.org/10.1017/pasa.2018.}7,  [arXiv: 1802.05076 [astro-ph.IM]].
%
\bibitem{burkespolaor}
S.Burke-Spolaor et al.,
Astron. Astrophys. Rev. {\bf 27} (2019) no. 5   [arXiv: 1811.08826 [astro-ph. HE]]. 

\doi{https://doi.org/10.1007/s00159-0190-0115-7}
%
%
%
\bibitem{AC}
T. Appelquist and J. Carazzone, 
\PRD{11,2856,1975}

\doi{https://doi.org/10.1103/PhysRevD.11.2856}
%
\bibitem{georgi}
H.M. Georgi, S.L. Glashow, M.E. Machcek and D. V. Nanopoulos, 
\PRL{40, 692,1978}

\doi{https://doi.org/10.1103/PhysRevLett.40.692}
%
\bibitem{inamikubotaokada}
T. Inami, T. Kubota and Y. Okada, 
Z. Phys. {\bf C 18}, 69 -80 (1983).

\doi{https://doi.org/10.1007/BF01571710}   
%
\bibitem{spira}
M. Spira, A. Djouadi, D. Graudenz and P.M. Zerwas, 
\NPB{453,17,1995}

\doi{https://doi.org/10.1016/0550-3213(95)00379-7}
%
%
\bibitem{wilczek}
F. Wilczek, 
\PRL{39,1304,1977}

\doi{https://doi.org/10.1103/PhysRevLett.39.1304}
%
\bibitem{shifman1}
M.A. Shifman, A.I. Vainstein, M.B. Voloshin, V.I. Zakharov, 
Sov. J. Nucl. Phys. {\bf 30}, 711 (1979); Yad. Fiz. {\bf 30}, 1368 (1979).
%
%
%
%
%
\bibitem{shifman2}
M.A. Shifman, A.I. Vainshtein, and V.I. Zakharov, 
\PLB{78,443,1978}

\doi{https://doi.org/10.1016/0370-2693(78)90481-1}
%
\bibitem{inamikubota}
T. Inami and T. Kubota, 
\NPB{179,171,1981}

\doi{https://doi.org/10.1016/0550-3213(81)90253-4}
%
%
\bibitem{sakai}
N. Sakai, 
\PRD{22,2220, 1980}

\doi{https://doi.org/10.1103/PhysRevD.22.2220}
%
\bibitem{braatenleveille}
E. Braaten and J.P. Leveille,  
\PRD{22,715,1980}

\doi{https://doi.org/10.1103/PhysRevD.22.715}

%
\bibitem{gaillardlee}
M.K. Gaillard and B.W. Lee, 
\PRD{10,897,1074}

\doi{https://doi.org/10.1103/PhysRevD.10.897}
%
%
\bibitem{inamilim}
T. Inami and C.S. Lim,
\PTP{65,297,1981}

\doi{https://doi.org/10.1143/PTP.65.297}
%
%
\bibitem{buras}
A.J. Buras,  
\PRL{46,1354,1981}

\doi{https://doi.org/10.1103/PhysRevLett.46.1354}
%
%
\bibitem{deshpande1}
N.G. Deshpande and G. Eilam, 
\PRD{26,2463,1982}

\doi{https://doi.org/10.1103/PhysRevD.26.2463}
%
%
%
%
\bibitem{deshpande2}
N.G. Deshpande and M. Nazerimonfared, 
\NPB{213,390,1983}

\doi{https://doi.org/10.1016/0550-3213(83)90228-6}
%
%
%
%
\bibitem{botella}
F.J. Botella and C.S. Lim, 
\PRD{34,301,1986}

\doi{https://doi.org/10.1103/PhysRevD.34.301}

%
%
%
\bibitem{degrasse}
G. Degrassi, E. Gabrielli and L. Trentadue, 
\PRD{79,053004,2009}
   [arXiv: 0812.3262 [hep-ph]].

\doi{https://doi.org/10.1103/PhysRevD.79.053004}

%
\bibitem{coriano1}
C. Corian{\` o}, L.Delle Rose, E. Gabrielli and L. Trentadue, 
\PRD{88,085008,2013}
   [arxiv: 1303.1305 [hep-th]].

\doi{https://doi.org/10.1103/PhysRevD.88.085008}
%
\bibitem{coriano2}
C. Corian{\` o}, L.Delle Rose, E. Gabrielli and L. Trentadue, 
\JHEP{03,136,2014}
 [arXiv: 1312.7657 [hep-ph]].

\doi{https://doi.org/10.1007/JHEP03(2014)136}
%
%
%
%
\bibitem{delbourgosalam}
R. Delbourgo and A Salam, 
%
``PCAC Anomalies and Gravitation"  

ICTP-preprint IC/72/86 (August,  1972).
%
\bibitem{delbourgosalam2}
R. Delbourgo and A Salam, 
\PLB{40,381,1972}

\doi{https://doi.org/10.1016/0370-2693(72)90825-8}
%
\bibitem{kimura}
T. Kimura,
\PTP{42,1191,1969}

\doi{https://doi.org/10.1143/PTP.42.1191}
%
\bibitem{eguchifreund}
T. Eguchi and P.G.O. Freund, 
\PRL{37,1251,1976}

\doi{https://doi.org/10.1103/PhysRevLett.37.1251}
%
\bibitem{ishamsalamstrathdee}
C.J. Isham, A. Salam and J. Strathdee, 
\PRD{3,1805,1971}

\doi{https://doi.org/10.1103/PhysRevD.3.1805}
%
%
\bibitem{kamefuchi1}
S. Kamefuchi, L. O'Raifeartaigh and A. Salam,
\NP{28,529,1961}

\doi{https://doi.org/10.1016/0029-5582(61)90056-6,10.1016/0029-5582(61)91075-6}
%
\bibitem{chisholm}
J.S.R. Chisholm, 
\NP{26,469,1961}

\doi{https://doi.org/10.1016/0029-5582(61)90106-7}
%
\bibitem{borchers}
H.-J. Borchers, 
\NC{15,784,1960}

\doi{https://doi.org/10.1007/BF02732693}
%
%
%


%
%

%
%

%
%
%
%

%
%
%

%
\bibitem{pdg}
P.A. Zyla {\it et al.}  (Particle Data Group), 
Prog. Theor. Exp. Phys. {\bf 2020}, 083C01 (2020). 

\doi{https://doi.org/10.1093/ptep/ptaa104}
%
\bibitem{eilam}
G. Eilam, 
\PRL{49,1478,1982}

\doi{https://doi.org/10.1103/PhysRevLett.49.1478}

\bibitem{hou}
W.S. Hou, 
\NPB{308,561,1988}

\doi{https://doi.org/10.1016/0550-3213(88)90578-0}
%
\bibitem{deshpande3}
N.G. Deshpande and J. Trampetic, 
\PRD{41,895,1990}

\doi{https://doi.org/10.1103/PhysRevD.41.895}
%
%
\bibitem{ruhdorfer}
M. Ruhdorfer, J. Serrs and A. Weiler, 
\JHEP{05,083,2020}

\doi{https://doi.org/10.1007/JHEP05(2020)083}
%



\end{thebibliography}

\end{document}